\documentclass[11pt,a4paper]{article}
\pdfoutput=1
\usepackage{jheppub}
\usepackage[utf8]{inputenc}
\usepackage{soul,xcolor}
\usepackage{cancel}
\usepackage{makecell}
\usepackage{diagbox}
\usepackage{makecell}
\usepackage{afterpage}
\newsavebox{\foobox}

\usepackage{multirow}
\usepackage{dcolumn}
\newcommand\Tstrut{\rule{0pt}{2.6ex}}         

\usepackage{ulem}
\usepackage{graphicx}
\usepackage{comment}
\usepackage{bm,array}
\usepackage{graphics}
\usepackage{epsf}
\usepackage{color}
\usepackage{amsmath}
\usepackage{amssymb}
\usepackage{latexsym}
\usepackage{slashed}
\usepackage{float}
\usepackage{cases}
\usepackage{multirow}
\usepackage{framed}
\def\be{\begin{equation}}
\def\ee{\end{equation}}
\def\ba{\begin{array}}
\def\ea{\end{array}}

\def\alambda{A_\lambda}
\def\akappa{A_\kappa}
\def\mueff{\mu_\mathrm{eff}}
\def\tanb{\tan\beta}

\def\wpm{W^\pm}
\def\wmp{W^\mp}
\def\sQ3{\widetilde{Q}_3}
\def\sU3{\widetilde{U}_3}
\def\sD3{\widetilde{D}_3}
\def\hone{h_1}
\def\htwo{h_2}
\def\aone{a_1}
\def\hsm{h_{\rm SM}}

\def\ntrli{\chi_i^0}

\def\ntrlone{\chi_1^0}
\def\ntrltwo{\chi_2^0}
\def\ntrlthree{\chi_3^0}
\def\ntrlfour{\chi_4^0}
\def\ntrlfive{\chi_5^0}

\def\ntrltwothree{\chi_{2,3}^0}
\def\ntrlthreefour{\chi_{3,4}^0}
\def\ntrltwothreefour{\chi_{2,3,4}^0}

\def\charonepm{\chi_1^\pm}

\def\mone{M_1}
\def\mtwo{M_2}

\def\mthree{M_3}
\def\mntrli{m_{{_{\chi}}_i^0}}

\def\mntrlone{m_{{_{\chi}}_{_1}^0}}
\def\mntrltwo{m_{{_{\chi}}_2^0}}
\def\mntrlthree{m_{{_{\chi}}_3^0}}
\def\mntrltwothree{m_{{_{\chi}}_{_{2,3}}^0}}

\def\mcharone{m_{{_{\chi}}_{_1}^\pm}}
\def\mchartwo{m_{{_{\chi}}_{_2}^\pm}}
\def\mhone{m_{h_1}}
\def\mhtwo{m_{h_2}}

\def\maone{m_{a_1}}

\def\vev{{\it vev}}

\def\vu{v_u}
\def\vd{v_d}

\def\vs{v_{_S}}
\newcommand{\fbinv}{\text{fb}$^{-1}$}

\def\etmiss{\slashed{E}_T}

\def\nmssmtools{{\tt NMSSMTools}}
\def\checkmate{{\tt CheckMATE~}}
\def\micromegas{{\tt micrOMEGAs}}
\def\higgsbounds{{\tt HiggsBounds}}
\def\higgssignals{{\tt HiggsSignals}}

\def\madgraph{{\tt MadGraph}}
\newcommand{\beq}{\begin{equation}}
\newcommand{\eeq}{\end{equation}}
\newcommand{\bea}{\begin{eqnarray}}
\newcommand{\eea}{\end{eqnarray}}
\title{Revisiting singlino dark matter of the natural $Z_3$-symmetric NMSSM
in the light of LHC}
\author[a,b]{Waleed Abdallah}
\author[c]{Arindam Chatterjee}
\author[a]{AseshKrishna Datta}
\affiliation[a]{Harish-Chandra Research Institute, HBNI, Allahabad 211019,
India}
\affiliation[b]{Department of Mathematics, Faculty of Science, Cairo University,
Giza 12613, Egypt}
\affiliation[c]{Physics and Applied Mathematics Unit, Indian Statistical
Institute, \\ 203 B.T. Road, Kolkata 700108, India}
\emailAdd{waleedabdallah@hri.res.in,
          arin\_t@isical.ac.in, asesh@hri.res.in}
\abstract{Inspired by the fact that relatively small values of the effective
higgsino mass parameter of the $Z_3$-symmetric Next-to-Minimal Supersymmetric
Standard Model (NMSSM) could render the scenario `natural', we explore the
plausibility of having relatively light neutralinos and charginos
(the electroweakinos or the ewinos) in such a scenario with a rather light singlino-like Lightest
Supersymmetric Particle (LSP), which is a Dark Matter (DM) candidate, and
singlet-dominated scalar excitations. By first confirming the indications in
the existing literature that finding simultaneous compliance with results from
the Large Hadron Collider (LHC) and those from various DM experiments with such
light states is, in general, a difficult ask, we proceed to demonstrate, with
the help of a few representative benchmark points, how exactly and to what
extent could such a highly motivated `natural' setup with a singlino-like DM
candidate still remains plausible.}
\keywords{Supersymmetry Phenomenology}
\begin{document}
\maketitle
%
%
\section{Introduction}
\label{sec:intro}
A key ingredient that renders a popular supersymmetry (SUSY) scenario like the
phenomenological Minimal or Next-to-Minimal Supersymmetric Standard Model
(pMSSM or pNMSSM) `natural'~\cite{Barbieri:1987fn, Ellis:1986yg, Giudice:2013nak} is a relatively small
SUSY conserving higgsino mass parameter `$\mu$' in the pMSSM~\cite{Baer:2012up, Baer:2013ava, Mustafayev:2014lqa, Baer:2015rja} or,
similarly, $\mueff$ in the pNMSSM. In both scenarios, this would imply presence
of at least two light neutralinos and a similarly light chargino
(electroweakinos or ewinos) which are higgsino-like. 

Though theoretically much motivated, such light ewinos generally derive
significant constraints from their null searches at the colliders. These
searches target pair or associated productions of such ewinos. A stronger set
of bounds emerge in scenarios with significant mass-splits between such states
and the lightest neutralino which is the Lightest SUSY Particle (LSP). The LSP
is stable when a well-known discrete symmetry called $R$-parity is conserved and
thus, can be a viable candidate for the Dark Matter (DM)~\cite{Goldberg:1983nd, Ellis:1983ew}.

The usual decay modes of the charginos and the neutralinos, when their spectrum
is not critically compressed and the squarks and the sleptons are much heavier,
involve on/off-shell gauge and Higgs boson(s) and are as follows:
\[ \charonepm \to \ntrlone \wpm{^{(*)}}, \:\: \ntrli \to \ntrlone
Z^{(*)}/h^{(*)}/a^{(*)} , \:\: \ntrli \to \charonepm \wmp{^{(*)}} \,, \quad (i=2,3,4,5) \]
where $h \, (a)$ is the scalar (pseudoscalar) Higgs boson. Then, the most
stringent constraints on `$\mu$' or $\mueff$ usually come from the studies of
associated $\charonepm \ntrltwothree$ productions with
$\charonepm \to \ntrlone \wpm{^{(*)}}$ and $\ntrltwothree \to \ntrlone Z$
leading to rather clean multi-lepton (up to 3 leptons) final states~\cite{Aaboud:2018jiw, Aaboud:2018sua, Sirunyan:2017lae,
Sirunyan:2018ubx}.\footnote{There have also been experimental searches
involving two
soft leptons~\cite{Sirunyan:2017lae, Aaboud:2017leg}, opposite sign di-leptons,
as well as final states with $b$-jets and photons~\cite{Sirunyan:2018ubx},
effectively constraining the chargino-neutralino spectra. The implications of
these searches for our present study will be discussed in some detail later in this work.}
Clearly, presence of a light enough Higgs boson could lead to a sizable
branching fraction for $\ntrltwothree \to \ntrlone h/a$ thus depleting the
lepton-rich events. This can potentially weaken the limit on `$\mu$' or $\mueff$~\cite{Ellwanger:2018zxt, Domingo:2018ykx} thereby opening up the parameter space
favored by `naturalness'.

In the MSSM, an optimally healthy split between $\ntrltwothree/\charonepm$ and
$\ntrlone$ is not possible when $\mu \ll \mone, \mtwo$ for which these states
are almost purely higgsinos and hence nearly degenerate, where $\mone$ and
$\mtwo$ stand for the soft SUSY-breaking masses of the $U(1)$ and $SU(2)$ gauginos, respectively. However, with $\mone < \mu < \mtwo$, one could find
$\mntrlone \sim \mone$, $\mntrltwothree \sim \mu \sim \mcharone $
and hence would obtain reasonable mass-splits
$\Delta m_{(\ntrltwothree, \ntrlone)}$ and $\Delta m_{(\charonepm, \ntrlone)}$
leading to hard enough leptons/jets in the cascades. This renders these
searches viable and hence yielding constraints.\footnote{For $\mtwo < \mu$, $\charonepm$ becomes
wino-dominated and degenerate in mass with a wino-dominated neutralino LSP. This
would result in softer leptons/jets in the cascades of $\charonepm$ thus eroding
experimental sensitivity to multi-lepton/jets final states, in general, and
tri-lepton final state, in particular~\cite{Abdallah:2017bkb}, while soft-lepton searches could emerge more relevant~\cite{Sirunyan:2017lae, Aaboud:2017leg}.} Critical studies as to how strong and
robust a constraint the LHC experiments could impose on such relatively light
higgsino-like ewinos (and hence on `$\mu$') have recently been undertaken by
various groups~\cite{Athron:2018vxy, Datta:2018lup}. Incidentally, the observed
SM-like Higgs boson ($\hsm$) of mass around 125~GeV could at best be the lightest of the
MSSM Higgs bosons while all its cousins have to be much
heavier (the so-called decoupling limit). Thus, there is only a limited scope
for the decay $\ntrltwothree \to \ntrlone \hsm$ to dominate over
$\ntrltwothree \to \ntrlone Z$ and hence weakening of lepton-rich final states
is not expected to be very common. Consequently, a notable relaxation on the
masses of these light ewinos (and hence on `$\mu$') is unlikely to be a
common occurrence.

In contrast, the situation can get very different in the NMSSM when the
coefficient `$\kappa$' of the superpotential term cubic in the singlet chiral
superfield gets vanishingly small (the Peccei-Quinn symmetric limit).
First, a rather light scalar ($\hone$) and a pseudoscalar ($\aone$) Higgs bosons with $m_{\hone, \aone} < m_Z$, both of which are singlet-dominated, are inevitable~\cite{Dermisek:2006wr, Huang:2013ima}. The Higgs sector of the
NMSSM has been studied in great details in refs.~\cite{Ellwanger:2009dp, Dermisek:2010mg, King:2012tr, Christensen:2013dra,
Cao:2013gba, King:2014xwa, Cao:2014kya, Bomark:2015hia, Bomark:2015fga,
Guchait:2015owa, Domingo:2015eea, Ellwanger:2015uaz, Costa:2015llh,
Conte:2016zjp, Das:2016eob,Baum:2017gbj, Ellwanger:2017skc,Baum:2019uzg}.
Second, a rather light singlino-dominated neutralino LSP (mass ranging from
sub-GeV to a few tens of a~GeV) is naturally present in the spectrum.
These two together could easily allow for a much smaller value of $\mueff$
leading to two next-to-LSP neutralino states ($\ntrltwothree$) and the lighter
chargino ($\charonepm$) all of which are higgsino-dominated with masses
$\sim \mueff$ and having prominent decays
$\ntrltwothree \to \ntrlone \hone/\aone$. Also, these facilitate the
simultaneous opening up of the decays $\ntrltwothree \to \ntrlone \htwo (\hsm)$
thus reinforcing the combined branching fractions of $\ntrltwothree$ to Higgs bosons over the same to $Z$-boson. Some specific consequences
of such possibilities had been studied in the past which include rather light
scalars decaying to (i) $\tau\bar{\tau}$ and leading to soft multi-lepton final 
state~\cite{Cerdeno:2013qta}, (ii) $b\bar{b}$~\cite{Dutta:2014hma} and
(iii) two photons~\cite{Ellwanger:2016wfe, Domingo:2016yih}. For
$\maone \lesssim 1$~GeV, even mesons can be produced out of a boosted pair
of light quarks that $\aone/\hone$ might decay to.\footnote{Furthermore, as we
would
appreciate later in this work, one could have a possible situation, without
sacrificing much of the essential features of such a scenario, when even the
decay $\charonepm \to \ntrltwo W^{\pm^*}$ could compete with
$\charonepm \to \ntrlone W^{\pm^*}$. The former would add to jet activity
via the decay $\ntrltwo \to \ntrlone \hone/\aone$ and hence could potentially
alter bounds obtained from the studies which vetoes extra jets. Otherwise,
BR($\charonepm \to \ntrlone W^{\pm^*}$) would remain 100\% and hence collider
constraints derived solely by studying $\charonepm$ pair production would hold
in a robust manner.} Recently, there has also been an attempt to an effective field theory approach and its connection to the NMSSM having Higgs portals in the description of thermal DM~\cite{Baum:2017enm}.

On the DM front, presence of singlet-like light scalars along with
a stable singlino-dominated LSP with a critical higgsino admixture (thanks to a
not so large $\mueff$), would have nontrivial consequences~\cite{Ellwanger:2016sur}. First, the higgsino admixture could now enable the LSP
annihilate efficiently enough in the early Universe yielding DM relic in the
right ballpark. Second, the same enhanced interaction of such an LSP could make
it sensitive to DM Direct Detection (DMDD) experiments. Third, 
the light scalars ($\aone$ and $\hone$) could offer new annihilation 
`funnels' that are efficient handles on the DM relic. Some aspects of such a singlino-higgsino mixed state has been discussed in ref.~\cite{Xiang:2016ndq} in reference to both as a DM candidate and its implications for the LHC. Furthermore, in the 
context of DMDD experiments sensitive to spin-independent (SI) 
scattering, there may appear the so-called blind spots~\cite{Cheung:2014lqa,Badziak:2015exr, Badziak:2016qwg, Badziak:2017uto, Cao:2018rix} either due
to vanishing LSP-Higgs coupling or due to a destructive interference between the
contributions from the $CP$-even Higgs bosons. These could suppress the 
DMDD-SI cross section to a value still allowed by experiments. 

The collider and the DM aspects of such an NMSSM scenario are thus expected to
be connected in a rather nontrivial way. 
It is encouraging to find a few recent works addressing 
these aspects, focussing mainly on one or the other of them. Their
broad scopes are as follows.  
\begin{itemize}
\item Ref.~\cite{Ellwanger:2018zxt} is the first one to discuss the case
of a light singlino-like LSP as the DM candidate with light bino (higgsino)-like 
neutralino(s) and a higgsino-like chargino as the next heavier
sparticle(s) and the combined constraint such a scenario draws from various DM
and collider experiments. It points out the roles played by relatively light
$\aone/\hone$ (i) in obtaining the DM Relic Density (DMRD) in the right ballpark,
(ii) in complying with the DMDD constraints on the SI scattering
cross section using the blind spot
mechanism and (iii) in evading (degrading) the LHC bounds on such light ewinos.
\item Ref.~\cite{Cao:2018rix} undertakes a detailed scan of the `natural'
NMSSM parameter space requiring relatively light higgsino-like states
compatible with the relic density bound from Planck experiment~\cite{Ade:2015xua, Aghanim:2018eyx}, the
bound from DMDD experiments like XENON-1T~\cite{Aprile:2018dbl, Aprile:2019dbj}
and those from the 13~TeV run
(with up to 36~\fbinv ~of data) of the LHC. In our current context, the most
relevant finding is that only a singlino-dominated LSP with a small
higgsino admixture ($\mueff \simeq \mntrlone$) might survive the combined
constraints if $\mntrlone \gtrsim 90$~GeV and, that also, for a compressed
spectrum for the LSP and $\charonepm$.
\item Ref.~\cite{Domingo:2018ykx} is mainly concerned with the impact of
recent multi-lepton searches at the LHC on the ewinos of the NMSSM in the
presence of light singlet scalars, $\hone$ and $\aone$. The study chooses to
remain agnostic about the detailed bounds in the DM sector except for
respecting only the upper bound on the relic density. The discussion on the
scenario with a relatively light singlino-like LSP that could annihilate via
singlet-Higgs funnel(s) are of particular relevance for our present work.
\end{itemize}

In this work we focus on an $Z_3$-symmetric NMSSM scenario with a relatively
small $\mueff$ (preferably less than $\sim 300$~GeV and not exceeding 500~GeV)
that ensures enhanced `naturalness' and with a singlino-enriched
($> 95\%$) LSP neutralino as the DM candidate with mass around or below
the SM Higgs boson funnel, i.e. $\lesssim 62$~GeV. The purpose is to 
find how such a scenario could still be compatible with all pertinent experimental data
from both DM and collider fronts. Our study goes beyond what was found in
ref.~\cite{Cao:2018rix} which excludes the possibilities of having a
singlino-dominated LSP below $\sim 90$~GeV and away from
the coannihilation regime. As we would elucidate soon,
allowing for some modest bino content in the lighter neutralinos by considering
an appropriately small $\mone$ could provide us with a much lighter and a 
viable singlino-dominated DM candidate which finds right funnels in various light
states like the SM Higgs boson, the $Z$-boson and even the lighter singlet-like Higgs states of the scenario to. This renders, not only the DM
neutralino, but the entire system of lighter neutralinos `well-tempered'~\cite{ArkaniHamed:2006mb}. 
Constraints imposed by us include the one on DMRD within 10\%
uncertainty, those from the DMDD experiments like XENON-1T
studying the SI~\cite{Aprile:2018dbl} and the spin-dependent
(SD)~\cite{Aprile:2019dbj,Amole:2019fdf} DM-nucleon scattering 
cross sections.
Our study also takes into account all relevant LHC analyzes that considers up to $\sim 36$~\fbinv 
~worth data via use of the package \checkmate\cite{Drees:2013wra, Dercks:2016npn}.

The paper is organized as follows. In section~\ref{sec:ewino-higgs} we present
the structures and the salient features of the NMSSM ewino and the Higgs sectors
along with their interactions that are relevant to the present work. These are
followed by a discussion on the nature of the spectrum in our scenario and on
the important decay modes of the light ewinos to Higgs bosons.
Section~\ref{sec:results} contains our results where the impact of experimental bounds
on the DM observables is quantitatively assessed leading to our choice of suitable
benchmark points with low enough $\mueff$. A dedicated {\tt CheckMATE}-based analysis follows to assess the viability of the benchmark points in view of the
LHC data. In section~\ref{sec:conclusions} we conclude.
%
\section{The light ewinos and the light Higgs bosons}
\label{sec:ewino-higgs}
%
The superpotential of the $Z_3$-symmetric NMSSM is given by
\beq
{\cal W}= {\cal W}_\mathrm{MSSM}|_{\mu=0} + \lambda \widehat{S}
\widehat{H}_u.\widehat{H}_d
        + {\kappa \over 3} \widehat{S}^3 \, ,
\label{eq:superpot}
\eeq
where ${\cal W}_\mathrm{MSSM}|_\mu=0$ is the MSSM superpotential sans the
higgsino mass term (the $\mu$-term), $\widehat{H}_u, \widehat{H}_d$ and
$\widehat{S}$ are the usual MSSM $SU(2)$ Higgs doublets and the
NMSSM-specific singlet superfields, respectively while `$\lambda$' and
`$\kappa$' are dimensionless coupling constants. The $\mu$-term is generated
dynamically from the second term when the singlet scalar field `$S$' develops
a vacuum expectation value (\vev) $\langle S \rangle$=$\vs$ (i.e., $\mueff=\lambda \vs$) thus offering
a solution to the puzzling $\mu$-problem~\cite{Kim:1983dt}.
The NMSSM-specific part of the soft SUSY-breaking Lagrangian is given by
\beq
-\mathcal{L}^{\rm soft}= -\mathcal{L_{\rm MSSM}^{\rm soft}}|_{B\mu=0}+ m_{S}^2
|S|^2 + (
\lambda A_{\lambda} S H_u\cdot H_d
+ \frac{\kappa}{3}  A_{\kappa} S^3 + {\rm h.c.}) \,,
\label{eq:lagrangian}
\eeq
where $m_S^2$ is the squared soft SUSY-breaking mass for the singlet scalar
field `$S$' while $\alambda$ and $\akappa$ are the NMSSM-specific trilinear soft
terms having dimensions of mass. In the following subsections we briefly
discuss the sectors that are directly involved in the present study, i.e., the
ewino and the Higgs sectors of the scenario.
%
\subsection{The ewino sector}
\label{subsec:ewinos}
The ewino sector is comprised of the neutralino and the chargino sectors.
The neutralino sector is augmented in the NMSSM by the presence of the singlino
($\tilde{S}$) state, when compared to the same for the MSSM. Thus, the symmetric
$5 \times 5$ neutralino mass matrix, in the basis
$\psi^0=\{\widetilde{B},~\widetilde{W}^0, ~\widetilde{H}_d^0, 
~\widetilde{H}_u^0, ~\widetilde{S}\}$ is given by~\cite{Ellwanger:2009dp}
\beq
\label{eq:mneut}
{\cal M}_0 =
\left( \begin{array}{ccccc}
\mone & 0 & -\dfrac{g_1 \vd}{\sqrt{2}} & \dfrac{g_1 \vu}{\sqrt{2}} & 0 \\[0.4cm]
& \mtwo & \dfrac{g_2 \vd}{\sqrt{2}} & -\dfrac{g_2 \vu}{\sqrt{2}} & 0 \\
& & 0 & -\mueff & -\lambda \vu \\
& & & 0 & -\lambda \vd \\
& & & & 2 \kappa \vs
\end{array} \right) \; ,
\eeq
where $g_1$ and $g_2$ stand for the gauge couplings of the $U(1)$ and $SU(2)$
gauge groups, respectively, and $\vu=v\sin\beta$, $\vd=v\cos\beta$ such that
$\tan\beta=\vu/ \vd$ with $v=\sqrt{\vu^2+\vd^2}~\simeq~174$~GeV.
%
The above mass-matrix can be diagonalized by a matrix $N$, i.e.,
\beq
N^* {\cal M}_0 N^\dagger 
= \mathrm{diag} (\ntrlone, \ntrltwo, \ntrlthree, \ntrlfour, \ntrlfive) \, .
\label{eq:diagN1}
\eeq
The resulting neutralino mass-eigenstates ($\ntrli$, in order of increasing mass as
`$i$' varies from 1 to 5), in terms of the weak eigenstates
($\psi_j^0$, with $j=1,\dots,5$), is given by
\beq
\chi_i^0 = N_{ij} \psi_j^0 \; .
\label{eq:diagN2}
\eeq
It is possible to find analytic expressions for the masses and the elements of the mixing matrix, $N_{ij}$, when two of the 
five states get decoupled. Hence, to start with, for our purposes, we consider the bino and the wino states to be decoupled. This would describe our 
basic setup fairly robustly with a rather light singlino-like LSP and a 
relatively small $\mueff$ (thus aiding `naturalness') leading to two light 
higgsino-like states. Such a scenario can be realized for
$\lambda v \ll |\mueff|$ along with $\kappa/\lambda \ll 1$. The ratios of 
higgsino to singlino admixtures in a given neutralino (in particular, in the
LSP) would remain to be much instrumental in our present analysis.
In the above-mentioned situation, these are given by~\cite{Cao:2015loa, Badziak:2015exr}
\beq
{N_{i3} \over N_{i5}}
  = {\lambda v \over \mueff} {(\mntrli/\mueff) \sin\beta - \cos\beta
                        \over {1 - (\mntrli/\mueff)^2}} \, ,
\quad \quad
{N_{i4} \over N_{i5}}
  = {\lambda v \over \mueff} {(\mntrli/\mueff) \cos\beta - \sin\beta
                        \over {1 - (\mntrli/\mueff)^2}} \, ,
\label{eqn:nij}
\eeq
where $N_{i3}$, $N_{i4}$ and $N_{i5}$ denote the two higgsino and the singlino
components, respectively, in the $i$-th mass eigenstate with $i=1,2,3$ and
$\mntrlone < \mntrltwo < \mntrlthree.$

Subsequently, we note that a relatively small value of $\mone$ ($\lesssim
\mueff$) could have a nontrivial impact on the combined DM and collider
phenomenology of such a scenario with light ewinos. However, this compels
one to work with a $4 \times 4$ neutralino
mass-matrix for which analytical expressions for $N_{ij}$, similar to those in
eq.~(\ref{eqn:nij}), would not be much illuminating. On top of that, when
$\mtwo$ is allowed to become small, the eigenvalue problem seeks solution of a
polynomial of degree 5 of which a general solution does not exist. Hence, for
smaller values of $\mone$ and/or $\mtwo$, we adopt a numerical approach.
On the other hand, the $2 \times 2$ chargino mass matrix of the NMSSM is
structurally the same as that of the MSSM with $\mu \to \mueff$ and,
in the basis
\beq
\psi^+ = \left( \begin{array}{c}
                 -i \widetilde{W}^+ \\
                 \widetilde{H}_u^+ 
                \end{array} \right) , \quad
\psi^- = \left( \begin{array}{c}
                 -i \widetilde{W}^- \\
                 \widetilde{H}_d^- 
                \end{array} \right) ,
\eeq
is given by~\cite{Ellwanger:2009dp}
\beq
{\cal M}_C = \left( \begin{array}{cc}
                    \mtwo   & \quad  g_2 \vu \\
                 g_2 \vd  & \quad \mueff 
             \end{array} \right) .
\eeq
As in the MSSM, this can be diagonalized by two $2 \times 2$ unitary matrices $U$ and $V$:
\beq
U^* {\cal M}_C V^\dagger = \mathrm{diag} (\mcharone , \mchartwo) \; ; \quad
\mathrm{with} \;\;  \mcharone < \mchartwo  \; .
\label{eq:uvmatrix}
\eeq
As noted in the Introduction, to ensure our scenario remains reasonably
`natural', we choose to work with relatively low values of $\mueff$. This yields
two light neutralinos along with a lighter chargino with masses $\sim \mueff$,
all of which can be dominantly higgsino-like. However, their actual masses and
compositions depend much on the extent they mix with the singlino and the bino
(for the neutralinos only) and with the wino states. In particular, we are
interested in a scenario where, $2 \kappa \vs \lesssim \mueff$ (i.e., for
$\kappa \lesssim {\lambda / 2}$). This could lead to a singlino-dominated
LSP. However, it may contain a crucial higgsino admixture thus
making it a viable DM candidate. Implications of such an LSP in the
context of various DM and collider experiments have recently been studied in the literature~\cite{Ellwanger:2018zxt, Cao:2018rix, Domingo:2018ykx}, though as
parts of more general studies.

Apart from the subtle role played by our proposed manoeuvring by allowing for
$\mone \lesssim \mueff$, this brings in a fourth relatively light neutralino in
the picture. We will further assume the wino-like neutralino to be the heaviest
of them all and hence would require $\mtwo > \mueff, \mone$. This would help
avoid stringent collider constraints by restricting heavier ewinos cascading via
such wino-like states. In the next subsection, we discuss that such a scenario
is necessarily accompanied by light singlet-like scalars which characterize our
scenario of interest.
%
\subsection{The Higgs sector}
\label{subsec:higgs}
%
The superpotential of eq.~(\ref{eq:superpot}) leads to the following
Lagrangian containing soft masses and couplings for the NMSSM Higgs sector:
\beq\label{2.5e}
-{\cal L}^\mathrm{soft} \supset
m_{H_u}^2 |H_u|^2 + m_{H_d}^2 | H_d |^2 
+ m_{S}^2 |S|^2
+\left(\lambda A_\lambda\, H_u \cdot H_d\; S + \frac{\kappa}{3}  A_\kappa\,
S^3 
+ \mathrm{h.c.}\right) .
\eeq
The neutral Higgs fields are parameterized about the real \vev's
$v_d$, $v_u$ and $\vs$ for the three neutral fields $H_d^0$,
$H_u^0$ and $S$, respectively as
\beq\label{2.10e}
H_d^0 = v_d + \frac{H_{dR} + iH_{dI}}{\sqrt{2}} , \quad
H_u^0 = v_u + \frac{H_{uR} + iH_{uI}}{\sqrt{2}} , \quad
S = \vs + \frac{S_R + iS_I}{\sqrt{2}},
\eeq
where ``$R$" and ``$I$" denote, for each field, the $CP$-even and the $CP$-odd
states, respectively. The $CP$-even squared mass matrix, ${\cal M}_S^2$, in the
basis $\{H_{dR}, H_{uR}, S_R\}$, is given by~\cite{Ellwanger:2009dp}
\beq\label{eqn:cp-even-matrix}
{\tiny{
{\cal
M}_S^2 =
\left( \begin{array}{ccc}
  g^2 v_d^2 + \mueff (\alambda+\kappa \vs) \,\tan\beta
& (2\lambda^2 - g^2) v_u v_d - \mueff (\alambda+\kappa \vs)
& \lambda (2 \mu_\mathrm{eff}\, v_d - (\alambda + 2\kappa \vs)v_u) \\[0.2cm]
  (2\lambda^2 - g^2) v_u v_d - \mueff (\alambda+\kappa \vs)
& g^2 v_u^2 + \mueff (\alambda+\kappa \vs) /\tan\beta
& \lambda (2 \mu_\mathrm{eff}\, v_u- (\alambda + 2\kappa \vs)v_d) \\[0.2cm]
  \lambda (2 \mu_\mathrm{eff}\, v_d - (\alambda + 2\kappa \vs )v_u)
& \lambda (2 \mu_\mathrm{eff}\, v_u - (\alambda + 2\kappa \vs)v_d)
& \lambda \alambda  \frac{v_u v_d}{\vs} + \kappa \vs (\akappa + 4\kappa \vs) 
\end{array} \right),}
}
\eeq
where $g^2=(g_1^2+g_2^2)/2$. The squared mass of the
singlet-like $CP$-even eigenstate (up to a mixing with the doublet states) is
given by the (3,3) component, i.e.,
\beq
{\cal{M}}^2_{S,33} =  
 \lambda \alambda  \frac{v_u v_d}{\vs} + \kappa \vs (\akappa + 4\kappa \vs) .
\label{eqn:cp-even-mass}
\eeq
Out of the other two eigenstates, one has to turn out to be the SM-like Higgs
boson with mass $\sim 125$~GeV, the other one being a relatively heavy,
doublet-dominated neutral Higgs boson with its squared mass around
$\mueff (\alambda + \kappa \vs)/\sin2\beta$. Thus, a more realistic basis to
work in is $\{H_1, H_2, S_R\}$, where $H_1=H_{dR}\cos\beta + H_{uR}\sin\beta$
and $H_2=H_{dR}\sin\beta - H_{uR}\cos\beta$, such $H_1$ resembles the SM Higgs
field.
Similarly, in the basis $\{A, S_I\}$, where $A=\cos\beta~H_{uI}+\sin\beta~H_{dI}$,
dropping the Goldstone mode, the $CP$-odd squared mass matrix ${\cal M}_{P}^2$
simplifies to~\cite{Ellwanger:2009dp}
\beq\label{eqn:cp-odd-matrix}
{\cal M}_P^2 =
\left( \begin{array}{cc}
    m_A^2
~&~ \lambda (\alambda - 2\kappa \vs)\, v \\[0.2cm]
    \lambda (\alambda - 2\kappa \vs)\, v
~&~ \lambda (\alambda + 4\kappa \vs)\frac{v_u v_d}{\vs} -3\kappa \akappa \, \vs  
\end{array} \right),
\eeq
with $m_A^2= 2 \mueff (\alambda + \kappa \vs)/\sin2\beta$ representing the
squared mass of the doublet-like $CP$-odd scalar, as in the MSSM. The mass-squared for the singlet
$CP$-odd scalar (modulo some mixing) is given by the (2,2) element of the
above matrix, i.e.,
\beq
{\cal{M}}^2_{P,22} =  
\lambda (\alambda + 4\kappa \vs)\frac{v_u v_d}{\vs} -3\kappa \akappa \, \vs \, .
\label{eqn:cp-odd-mass}
\eeq
The mass eigenstates of the $CP$-even ($h_i$) and the $CP$-odd ($a_i$) sectors
are given by~\cite{Badziak:2015exr, Badziak:2017uto}
\bea
h_i  &=& E_{h_i H_1} H_1 + E_{h_i H_2} H_2 + E_{h_i S_R} S_R, \quad (i=1,2,3) \\
[0.2cm]
a_i &=& O_{a_i A} A + O_{a_i S_I} S_I,  \quad (i=1,2)  
\eea
where $E$ ($3 \times 3$) and $O$ ($2 \times 2$) are the matrices that
diagonalize the mass-squared matrices for the $CP$-even scalars in the basis
$\{H_1, H_2, S_R\}$ and that for the $CP$-odd scalar of
eq.~(\ref{eqn:cp-odd-matrix}).

Clearly, the scalar masses have rather complex dependencies on as many as six
input parameters like $\lambda$, $\kappa$, $\alambda$, $\akappa$, $\mueff$ and
$\tan\beta$. However, for our scenario of interest with a light singlino-like
LSP ($\mntrlone \approx m_{\tilde{S}} \sim 2 \kappa \vs$) and light singlet-like
scalars (given by eqs.~(\ref{eqn:cp-even-mass}) and (\ref{eqn:cp-odd-mass})),
one could find the following (approximate) sum-rule~\cite{Das:2012rr, Ellwanger:2018zxt} relating their masses
when the singlet-doublet mixing among the scalar (Higgs) 
states can be safely ignored, i.e., in the decoupling limit
($\lambda, \; \kappa \to 0$) or for a sizable $\tan\beta$ and not too large $\lambda$, $\kappa$ and $\alambda$:
\beq
{\cal M}_{0,55}^2 \simeq {\cal M}_{S,33}^2 + {1 \over 3} {\cal M}_{P,22}^2
\quad \Rightarrow \quad \mntrlone^2 \simeq m_{h_1}^2
     + {1 \over 3} m_{a_1}^2 \, .
\eeq
This clearly indicates that the masses of the singlino and those for
the singlet-like scalar and the pseudoscalar are rather closely tied. The 
relationship becomes handy in discussions on DM-annihilation via light scalar funnels~\cite{Ellwanger:2018zxt, Cao:2018rix}.
%
\subsection{Interactions among the ewinos and the scalars}
\label{subsec:inter}
Interactions among the ewinos and the Higgs-like scalars states take the
central stage in our present study. Their subtle dependence on various NMSSM
parameters and their interplay crucially shape the phenomenology on both DM and
collider fronts, sometimes in a rather complementary fashion.

To be a little more specific, conformity with the observed value of DMRD 
would depend not only on a mass-spectrum that offers efficient
DM-annihilation mechanisms via funnels and/or coannihilations\footnote{As
mentioned in the Introduction, our focus would be on the region of the NMSSM
parameter space where funnel-assisted annihilation of DM occurs.
Note that we are interested in rather small values of 
$\mu_{\rm eff}$ in the case of an uncompressed spectra. This leads to rather 
light $\ntrlone$, and funnel assisted annihilation provides an opportunity 
to achieve the right thermal relic abundance in these regions of the 
parameter space.} but also on the strengths of the involved interactions. 
The latter, in turn, could also control the DM-nucleon interactions 
that are studied at the DMDD 
experiments. Hence requiring an efficient
DM-annihilation to meet the DMRD observations might imply a strong enough
DM-nucleon interaction strength that is ruled out by the DMDD experiments.
The converse is also true. This highlights a built-in tension in finding a
simultaneous explanation of the two crucial observations in the DM sector
alone.

On the collider front, the interactions among the ewinos and the
scalars determine the branching fractions of the former to the latter.
Such modes include the ones beyond what are being routinely considered in the
LHC analyzes in the context of `simplified scenarios' and result in new final
states. These are likely to result in relaxed mass-bounds on the higgsino-like
states thus offering enhanced `naturalness'. Interestingly enough, as we will
discuss soon in section~\ref{sec:results}, these might also help  satisfy the constraints in the DM sector.
Thus, for decoupled sfermions and a gluino, the interactions that are of
paramount importance are those among (i) various neutralinos and the gauge
($Z$-) boson and (ii) various neutralinos and Higgs bosons, of both
$CP$-even (scalar) and $CP$-odd (pseudoscalar) types, from both doublet and the singlet sectors.

The neutralino DM interacts with the $Z$-boson only through its higgsino
admixture. This interaction governs the self-annihilation of DM via $Z$-boson funnel
thus controlling the DMRD as well as the DMDD-SD cross section
and is given by~$\alpha_{Z\ntrlone\ntrlone}~\sim~|N_{13}^2-N_{14}^2|$~\cite{Haber:1984rc}.
On the other hand, a doublet-like Higgs scalar has an MSSM-like interaction with a higgsino and a gaugino. In addition, in the
$Z_3$-symmetric
NMSSM, as can be gleaned from eq.~(\ref{eq:superpot}), this also interacts with a higgsino and a singlino while the singlet-like scalar interacts with two higgsinos, both strengths being proportional to `$\lambda$'.
The $\widehat{S}^3$ term
in eq.~(\ref{eq:superpot}) further implies that the singlet scalar has an interaction with two singlinos whose strength goes as `$\kappa$'.
Thus, if the gaugino 
(bino and/or wino) admixture in a singlino-dominated LSP can be ignored 
(which is somewhat ensured by the neutralino mass matrix of the NMSSM), 
the generic coupling of such an LSP with the $CP$-even Higgs scalars are given by~\cite{Badziak:2015exr}
\bea
\alpha_{h_i \ntrlone \ntrlone}
 &\approx &
\sqrt{2} \lambda \Big[ E_{h_i H_1} N_{15} (N_{13} \sin\beta + N_{14} \cos\beta)
 + E_{h_i H_2} N_{15} (N_{14} \sin\beta - N_{13} \cos\beta)  \nonumber \\
 & & \hskip 30pt + \; E_{h_i S_R} (N_{13}N_{14} - {\kappa \over \lambda} N_{15}^2)
\Big].
\label{eqn:hn1n1}
\eea
The couplings $\alpha_{a_i \ntrlone \ntrlone}$ for the $CP$-odd scalar
counterparts
would be somewhat similar except for the appearance of an overall factor of 
imaginary `$i$' and that in the rotated basis for the pseudoscalar sector there
are only two massive eigenstates.

Furthermore, while the $CP$-even Higgs states from the doublet and the singlet 
sectors contribute to both DMRD and DMDD-SI, their $CP$-odd counterparts could 
contribute only to DMRD and practically nothing to any DMDD processes~\cite{Jungman:1995df}.
Ref.~\cite{Badziak:2015exr} discusses the issue of the blind
spots for DM-nucleon interaction in a few specific and motivated 
scenarios in  the $Z_3$-symmetric NMSSM. Among these, the scenario that
is germane to our present study is the one (section 6) that discusses blind
spots arising from destructive interference between $CP$-even singlet-like 
($\hone$) and doublet-like ($\htwo$, the observed SM-like Higgs boson) scalar 
states where $\mhone < \mhtwo$, while the heavier MSSM-like $CP$-even Higgs
state is virtually decoupled. This yields a DM-nucleon SI cross section well 
below the threshold of sensitivity of the relevant DMDD-SI experiments.
%
\begin{figure}[t]
\centering
\includegraphics[height=0.27\textheight, width=0.49\columnwidth ,
clip]{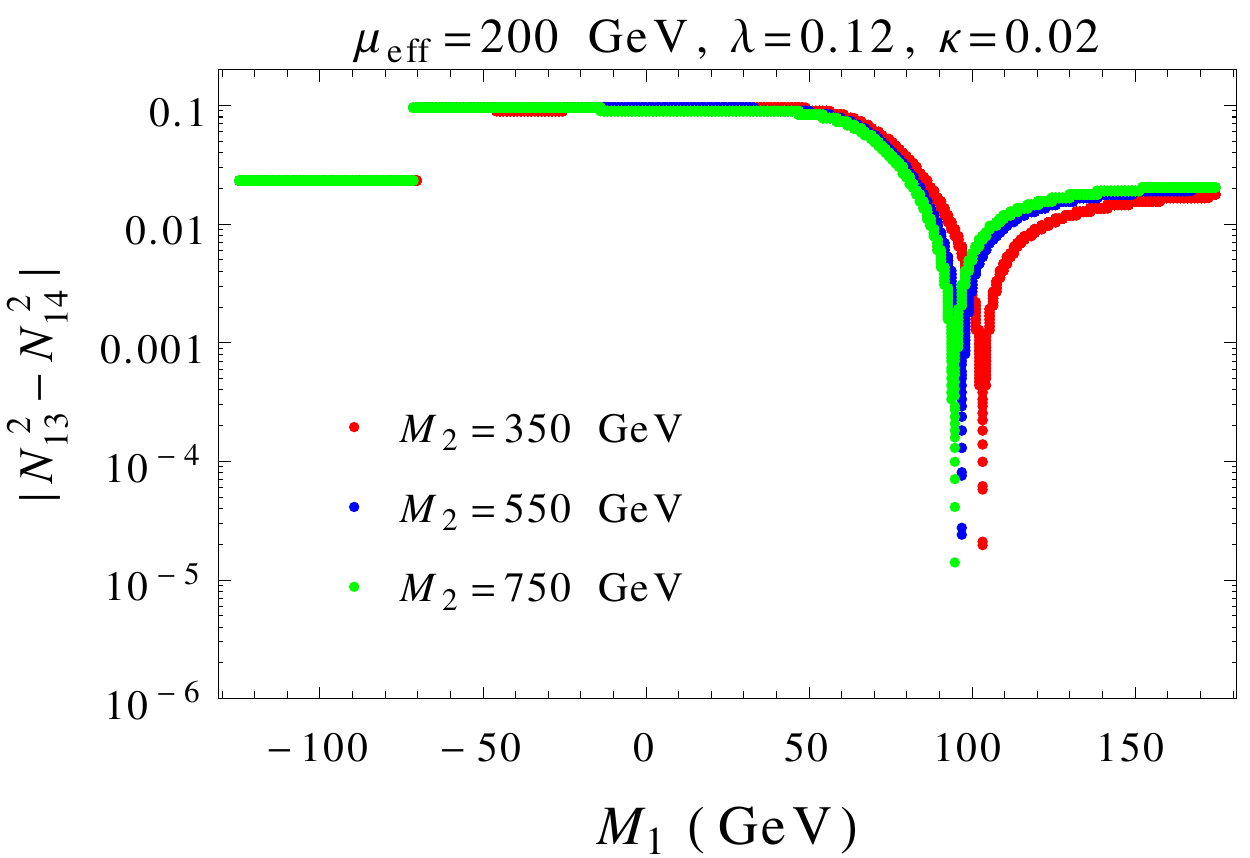}~~
\includegraphics[height=0.27\textheight, width=0.49\columnwidth ,
clip]{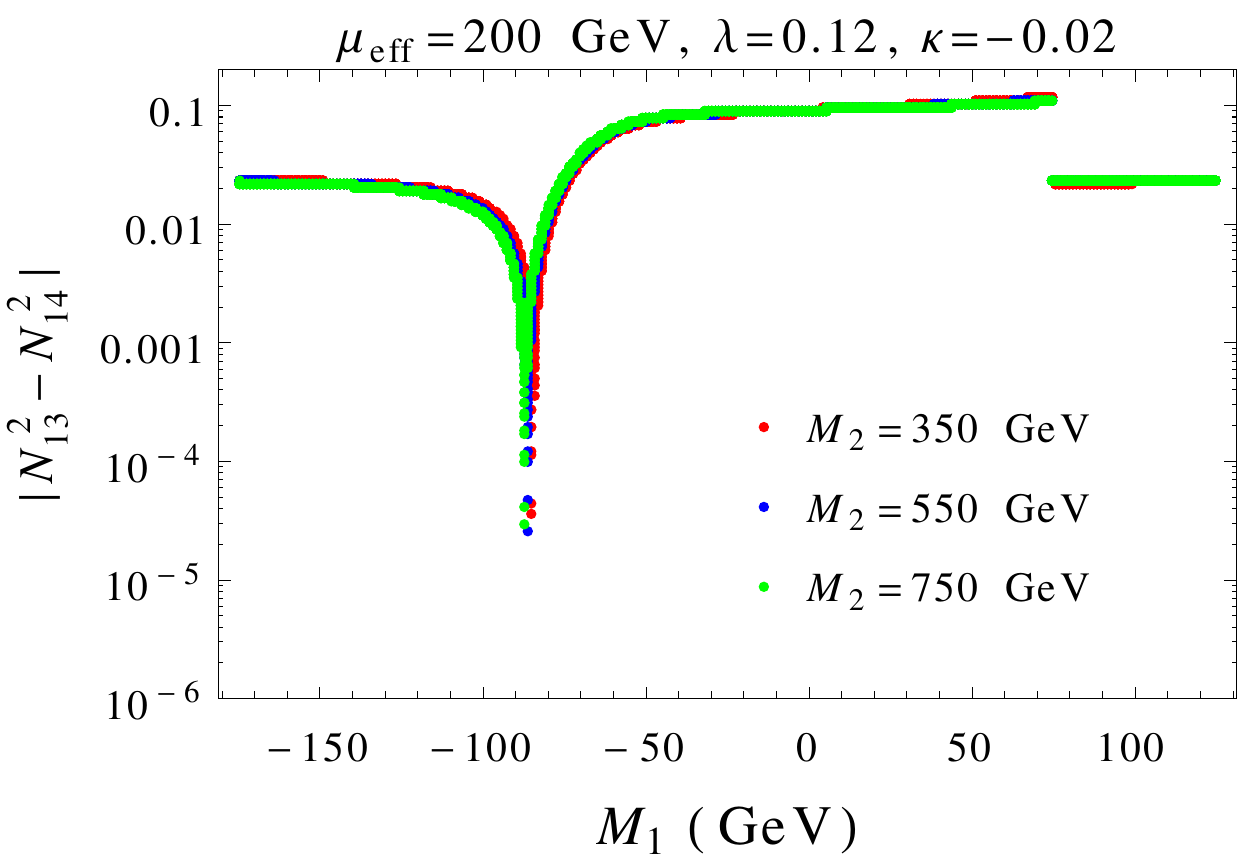}
\vskip 20pt
\includegraphics[height=0.29\textheight, width=0.49\columnwidth ,
clip]{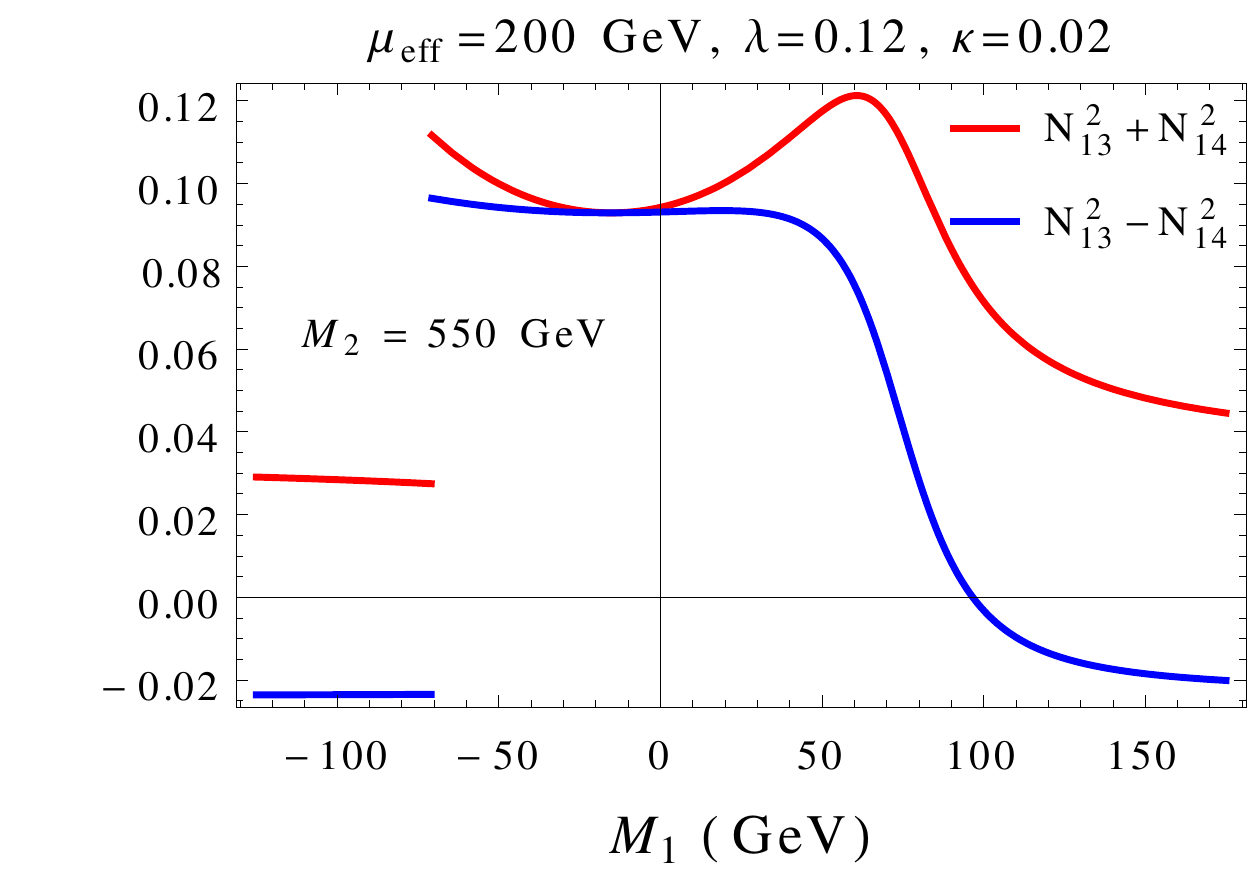}~~
\includegraphics[height=0.29\textheight, width=0.49\columnwidth ,
clip]{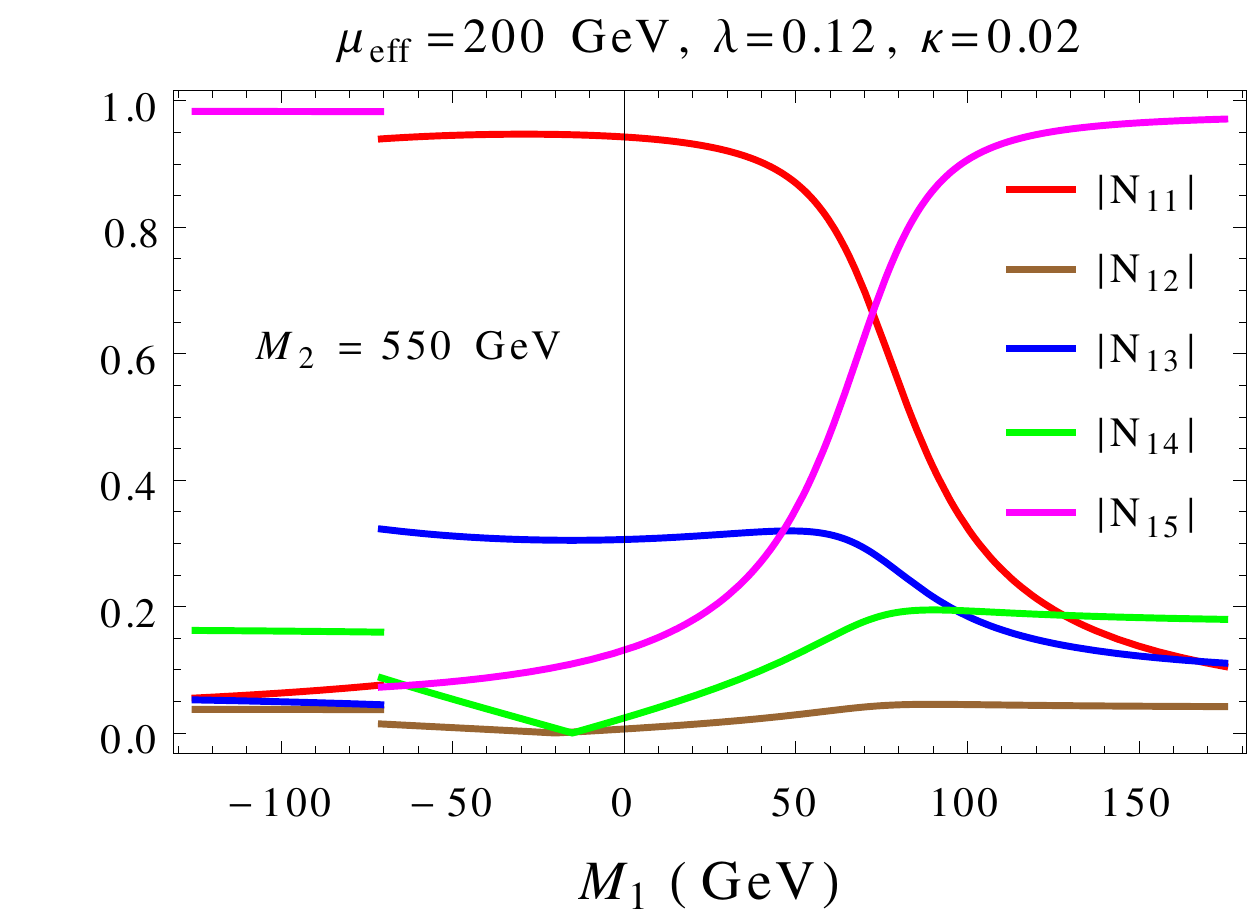}
\caption{{\bf Upper panel:} Variation of the quantity
$|N_{13}^2-N_{14}^2|$ as a 
function of $\mone$ for three values of $\mtwo$ (350~GeV, 550~GeV and 750~GeV)
for $\mueff=200$~GeV, $\lambda=0.12$, and for $\kappa > 0 \; (<0)$ for upper left (upper right)
plot. {\bf Lower panel:} Corroborative plots showing variations of the 
actuals (signed) values of the quantities $N_{13}^2-N_{14}^2$ and
$N_{13}^2+N_{14}^2$ (left) and $|N_{1j}|, \; (j=1, \dots, 5)$ (right) as 
functions of $\mone$ for a fixed value of $\mtwo \, (=550 \, \mathrm{GeV})$. $\tanb$ is set to 40 throughout. Values of fixed input parameters are
indicated at the top of each plot and are the same for all the plots. Plots
are created using \texttt{SARAH}-{\tt v4.9.0}-generated~\cite{Staub:2008uz, Staub:2013tta}
NMSSM model and hence present tree-level values only.}
\label{fig:m1-m2-var}
\end{figure}
%

Up to this point, the relative strengths of all the couplings that matter
are essentially governed  the ratios presented in eq.~(\ref{eqn:nij}).
It is now instructive to note that if the
singlino-dominated LSP could be infused with a bino/wino component,
it would alter the higgsino shares in the same.
This can be achieved by allowing bino/wino to mix
substantially with higgsinos, given that, at the lowest order, this only can
(indirectly) induce some gaugino admixture in an otherwise singlino-dominated
LSP. Hence such a regime would reign as long as $\mone$ (or $\mtwo$,
though decreasing it beyond a point could attract severe experimental constraints) is not too 
far away from $\mueff$. In certain regions of the NMSSM parameter space, with $m_{\tilde S} < \mone < \mueff$, this causes the coupling-strength 
$\alpha_{Z\ntrlone\ntrlone}~\left(\sim |N_{13}^2-N_{14}^2|\right)$
weakening to a minimum due to rather involved variations of $N_{ij}$'s as
functions of $\mone$. This we will discuss soon in a little more detail. This would then diminish the DMDD-SD cross section thus helping us evade the
related experimental bound. Clearly, under such circumstances,
eq.~(\ref{eqn:nij}) ceases to hold and improving
the same in the presence of an active bino state is unlikely to be illuminating
enough, given the complicated structure the situation presents. We thus take a numerical
route for the rest of the present study and frequently confront the results
with broad-based expectations for checking their basic sanity.

It may further be noted that the higgsino content of the LSP (given by $N_{13}^2+N_{14}^2$) could contribute only
partially to the DMDD-SI cross section (for the DM-nucleon scattering process mediated by the doublet $CP$-even Higgs 
bosons) while there could be a significant additional contribution from the singlet-like Higgs exchange in such a scattering. However, in the region of parameter space of our interest for which $\kappa \sim {\cal O} (10^{-2})$, this contribution is expected to be suppressed. 
An increase in the total higgsino fraction could attract severe experimental 
constraints from the DMDD-SI experiments. However, its effect
may get subdued in the presence of 
blind spots in the SI processes. In this work we exploit these two 
simultaneous effects in our favour, by manoeuvring $\mone$ and/or $\mtwo$, 
to find compliance with the DMDD data while still obtaining a
thermal relic density within the Planck-allowed range.

In the upper panel of figure~\ref{fig:m1-m2-var} we present the variations of the 
quantity $|N_{13}^2-N_{14}^2|$ as a function of 
$\mone$ (which can take both signs), for three different values of $\mtwo$
(350~GeV, 550~GeV and 750~GeV) with $\mueff=200$~GeV, $\tanb=40$ and with
$\kappa >0 \; (<0)$ on the upper left (upper right) plot. One clearly finds that 
the magnitude of $|N_{13}^2-N_{14}^2|$ could practically drop to a vanishing level as $\mone$ decreases.
However, as can be gleaned from the plots in the upper panel, for what 
exact value of $\mone$ this happens, depends on the input value of $\mtwo$,
although it becomes more or less insensitive to $\mtwo$ for its larger values.
These two plots also reveal that such a phenomenon occurs only for $\mone$ and 
`$\kappa$' having no relative sign between them, a situation in which the mixing 
between the two involved sectors is known to get maximal. It is worth pointing 
out that even though the relevant null entry in the neutralino mass matrix
(eq.~(\ref{eq:mneut})) prohibits a direct mixing between the bino and the 
singlino states, a possible mixing through the higgsino portal could give rise
to something that drastic with important phenomenological consequence akin to a 
blind spot for DMDD-SI scattering, but this time occurring for DMDD-SD 
scattering. We exploit this effect in our study the results of which are 
presented in section~\ref{sec:results}. The onset of discontinuous flat 
line segments seen at the top left (right) part of the plots on left (right)
has its origin in the bino-like LSP with a negative mass-eigenvalue turning instantly to a singlino-dominated one with a positive eigenvalue, for certain
particular values of $\mone$ depending on values of other input parameters.

Plots in the lower panel of figure~\ref{fig:m1-m2-var} explain the general 
behaviours of the ones in the upper panel by studying the variations of the 
components but referring only to the upper left plot having $\kappa >0$.
The left plot in the lower panel shows that the quantity $N_{13}^2-N_{14}^2$ 
indeed changes sign while varying smoothly, passing through a vanishing value,
as $\mone$ decreases. Over this region, the quantity $N_{13}^2+N_{14}^2$ (controlling the DMDD-SI rate) also grows smoothly with a decreasing $\mone$. Patterns of these variations find support in the individual variations of 
$|N_{13}|$ and $|N_{14}|$ with $\mone$ as illustrated in the bottom right plot
where, in particular, we see a cross-over point of the blue (representing 
$|N_{13}|$) and the green (representing $|N_{14}|$) curves thus explaining a 
vanishing value for $N_{13}^2-N_{14}^2$ seen in the lower left plot. It worths
a mention that the crucial variation is the one that of $|N_{13}|$. While it may 
not be outright unexpected that lowering of $\mone$ would immediately result in 
an enhanced bino admixture in the LSP, at the expense of mostly a decreasing 
singlino fraction in the same, it is somewhat curious to note that a decreasing 
$\mone$ boosts the otherwise subdominant higgsino content of the LSP in the form 
of $|N_{13}|$. It is possible that a decreasing $\mone$, given its healthy 
connection to the higgsino sector, drags the higgsino along on a collective bid 
to deplete the singlino content in the LSP. The discontinuity of the curves
appearing for certain negative $\mone$ values in the upper left plot are also
efficiently explained by the plots in the lower panels.
%
\subsection{The spectrum and the decays}
\label{subsec:spec-dk}
%
Discussions in the previous susbsections reveal that both the light
(singlet-like) Higgs sector and the neutralino sector get simultaneously
affected in a rather intricate way as `$\kappa$' turns smaller. This includes
non-trivial modifications of the involved couplings among these states via
mixings effects in both sectors and resulting mass-splits between the
physical states. Together these could alter the phenomenology in an essential
manner and experimental analyzes need to take due note of the same.

As has been already pointed out, in the scenario under study, the lightest
neutralino (the LSP) is singlino-like whereas the immediately heavier neutralinos, 
to start with, are higgsino-like. The latter could have enhanced
decay branching fractions to singlet-like Higgs bosons, $\hone$ and $\aone$,
which can become light enough for suitably small values of `$\kappa$'. Under
such a circumstance, the SM-like Higgs boson is the second lightest $CP$-even
Higgs boson ($\htwo \sim \hsm$) and this is always the case in our present study. 
As mentioned in the Introduction, the decay branching fractions of
the neutralinos to lighter (singlet-like) Higgs bosons could then compete with
(or could even exceed) those for the popularly considered modes like
$\ntrltwothree \to \ntrlone Z^{(*)}/\htwo (\hsm)$ and this is likely to relax the
existing bounds on the ewino sector.

A further nontrivial alteration of the decay branching fractions of the
higgsino-like neutralinos may take place if one allows for a bino/wino-like 
neutralino (now $\ntrltwo$) sneak in below the formerly higgsino-like states
(now $\ntrlthreefour$). Thus, more involved cascades could kick in, viz.,
$\ntrlthreefour \to \ntrltwo(\to\ntrlone h_i ) Z/h_i$ (with $i=1,2,3$ standing for the two light singlet-like and the SM-like Higgs bosons)
thanks to some
higgsino admixture in an otherwise gaugino (bino)-dominated $\ntrltwo$. This would have important bearing on
collider phenomenology. In section~\ref{sec:results}, we shall discuss how such
an intermediate state plays a crucial role in finding an all-round compliance with the experimental results pertaining to the DM-sector (as pointed out in
section~\ref{subsec:inter}) as those from the colliders. 

In passing, it is to be noted that presence of light Higgs states would
not directly affect the decay of the lighter chargino for which the experiments
assume BR($\charonepm \to \ntrlone \wpm{^{(*)}}$) to be 100\% when the
other Higgs states of the NMSSM, along with the sfermions, are all much heavier. Thus, at the first sight, it might appear that bounds imposed on the lighter
chargino sector, in particular, by looking for its pair production, and,
consequently, on $\mueff$ (for a higgsino-like lighter chargino) would still hold and need to be respected.
However, there are a couple of caveats. First, since the presence of a light
singlino state could significantly modify the NMSSM neutralino spectrum through
its mixing with the light higgsino states, a reasonable mass-split between
$\ntrltwothree$ and $\charonepm$ cannot be ruled out. This could open up competing decay modes of $\charonepm$ in the form
$\charonepm \to \ntrltwothree \wpm{^{(*)}}$. While these would still lead
to final states with leptons thanks to the presence of $\wpm{^{(*)}}$, the same are
likely to be contaminated with the decay products of $\ntrltwothree$, as noted
in the last paragraph. Second, as discussed above in the case for the
neutralinos, the competing decay mode in the form of
$\charonepm \to \ntrltwo \wpm{^{(*)}}$ could again open up for the lighter chargino when we require, as discussed in section~\ref{subsec:inter}, $\mone$ to be brought down below $\mueff$.
In both cases, experimental bounds even from the study of chargino pair production would likely to get relaxed.\footnote{Note, however, that if
$\mtwo \lesssim \mueff$, this would present us with a lighter chargino which is 
wino-like and close in mass with $\ntrltwo$. Hence the second effect mentioned
above would be absent and BR$(\charonepm \to  \ntrlone W^\pm)$ would
be 100\%. This would thus invite the standard, stronger bound on $\mtwo$ from
null searches for chargino pair production at the LHC.}

As for the light Higgs states ($\hone, \, \aone$) appearing in the cascades of
the lighter neutralinos, those could have significant branching fractions to
$b \bar{b}$ similar to the case of the SM-like Higgs boson. However, in general,
constraints derived from neutralino cascades involving such Higgs states are
weaker when compared to those obtained with cascades involving
$Z^{(*)}$~\cite{Sirunyan:2018ubx}.
Thus, enhanced branching fraction for the decay $\ntrltwothree \to \ntrlone
\hone/\aone$ (at the expense of BR($\ntrltwothree\!\to~\!\!\ntrlone Z^{(*)})$) are
expected to relax the existing experimental bounds on the ewino sector thus
capable of opening up a more `natural' region of the NMSSM parameter space.

In this work we confine ourselves to a region of the $Z_3$-symmetric NMSSM
parameter space for which the LSP is a singlino-dominated ($> 95\%$), 
the lighter chargino and two neutralinos are higgsino-like with
masses $\lesssim 300$~GeV, with a further possibility of having an intermediate
(gaugino-like) neutralino lighter than the higgsino-like states. In addition,
the setup offers singlet-like scalars that are lighter than the SM-like Higgs
bosons which could even turn out to be lighter than the LSP. Phenomenological
possibilities discussed in the previous paragraphs are realized in such a set
up. Scan-ranges adopted for various model parameters are summarized in table~\ref{tab:ranges}.
The soft masses for the $SU(3)$ gaugino ($M_3$), those for the sfermions
and the soft trilinear parameters $A_{\tau,b,t}$ are all fixed at around 5~TeV
while $A_{e,\mu}$ is set to zero.
\begin{table}[t]
\begin{center}
\begin{tabular}{|c|c|c|c|c|c|c|c|c|}
\hline
\makecell{Varying \\ parameters}  & $\lambda$ & $|\kappa|$ & $\tanb$& \makecell{$|\mueff|$ \\ (GeV)}&  \makecell{$|\alambda|$ \\ (TeV)} &
\makecell{$|\akappa|$ \\ (GeV)}
 & \makecell{$\mone$ \\ (GeV)} & \makecell{$\mtwo$ \\ (TeV)} \\
\hline
  & 0.05--0.2 & 0.001--0.05&
1--60& $\leq 300$ & $\leq 10$ & $\leq 100$ & 50--500& 0.2--1\\
\hline
\end{tabular}
\caption{Ranges of various model  parameters adopted for
scanning the $Z_3$-symmetric NMSSM parameter space. All parameters are defined at the scale
$Q^2 = (2m_{\tilde{Q}}^2 + m_{\tilde{U}}^2+m_{\tilde{D}}^2)/4$, except for $\tanb$ which is defined at $m_Z$ (see text for details).}
\label{tab:ranges}
\end{center}
\end{table} 
%
\section{Results}
\label{sec:results}
%
We now present our results for the broad scenario discussed in the previous
section which is characterized by a light
singlino LSP, accompanied by rather light singlet-like scalars, along with
higgsino-dominated lighter chargino and neutralinos that ensure a healthy
degree of `naturalness'. 
The focus is on if such a scenario can be compatible with
recent constraints pertaining to the DM sector (i.e., those involving DMRD,
DMDD-SI and DMDD-SD) and those coming from various past and recent collider
experiments that include the LEP and the LHC experiments. In particular, it
emerges from the recent literature~\cite{Cao:2018rix, Domingo:2018ykx}
that such an all-round compliance is not easy to find. As pointed out in the
Introduction, ref.~\cite{Cao:2018rix} concludes that this may be only
possible in the
coannihilation region marked by a near-degeneracy of the singlino-dominated LSP and the higgsino-dominated chargino (and neutralinos). Our goal is to go beyond
this and to find if such a thorough compliance with DM and collider data is
possible away from the coannihilation region while still retaining the
essential features of the broad scenario.

Results are obtained via a random scan over the parameter space of the
$Z_3$-symmetric NMSSM using the package \nmssmtools-{\tt v5.1.0}~\cite{Ellwanger:2004xm, Ellwanger:2005dv, Das:2011dg}.
Experimental constraints (at $2\sigma$ level) implemented in \nmssmtools~ are
automatically imposed on our analysis. These include various constraints
from the LEP experiments, including the one pertaining to invisible decay width of the $Z$-boson, and those on the $B$-physics observables.
Compliance with experimental results on $(g-2)_\mu$ is not demanded. In addition, constraints
from various Higgs boson searches at LEP and Tevatron and compatibility to the
Higgs boson observed at the LHC are considered/checked by using the packages
\higgsbounds-{\tt v4.3.1}~\cite{Bechtle:2008jh, Bechtle:2011sb} and
\higgssignals-{\tt v1.4.0}~\cite{Bechtle:2013xfa, Bechtle:2014ewa}, which, among other things, ensures compliance with the upper bound on the invisible decay width of the observed Higgs boson.  
DM-related computations are done
using an adapted version of the package \micromegas-{\tt v4.3}
\cite{Belanger:2006is, Belanger:2008sj, Barducci:2016pcb} that is
built-in to \nmssmtools. Finally, we employ the package
\texttt{CheckMATE}-{\tt v2.0.26}~\cite{Drees:2013wra, Dercks:2016npn} to check
our benchmark points (that pass all relevant constraints including
the DM-related ones) if they are passing all relevant LHC analyzes.
%
%
\begin{figure}[t]
\centering 
\includegraphics[height=0.3\textheight, width=0.6\columnwidth ,
clip]{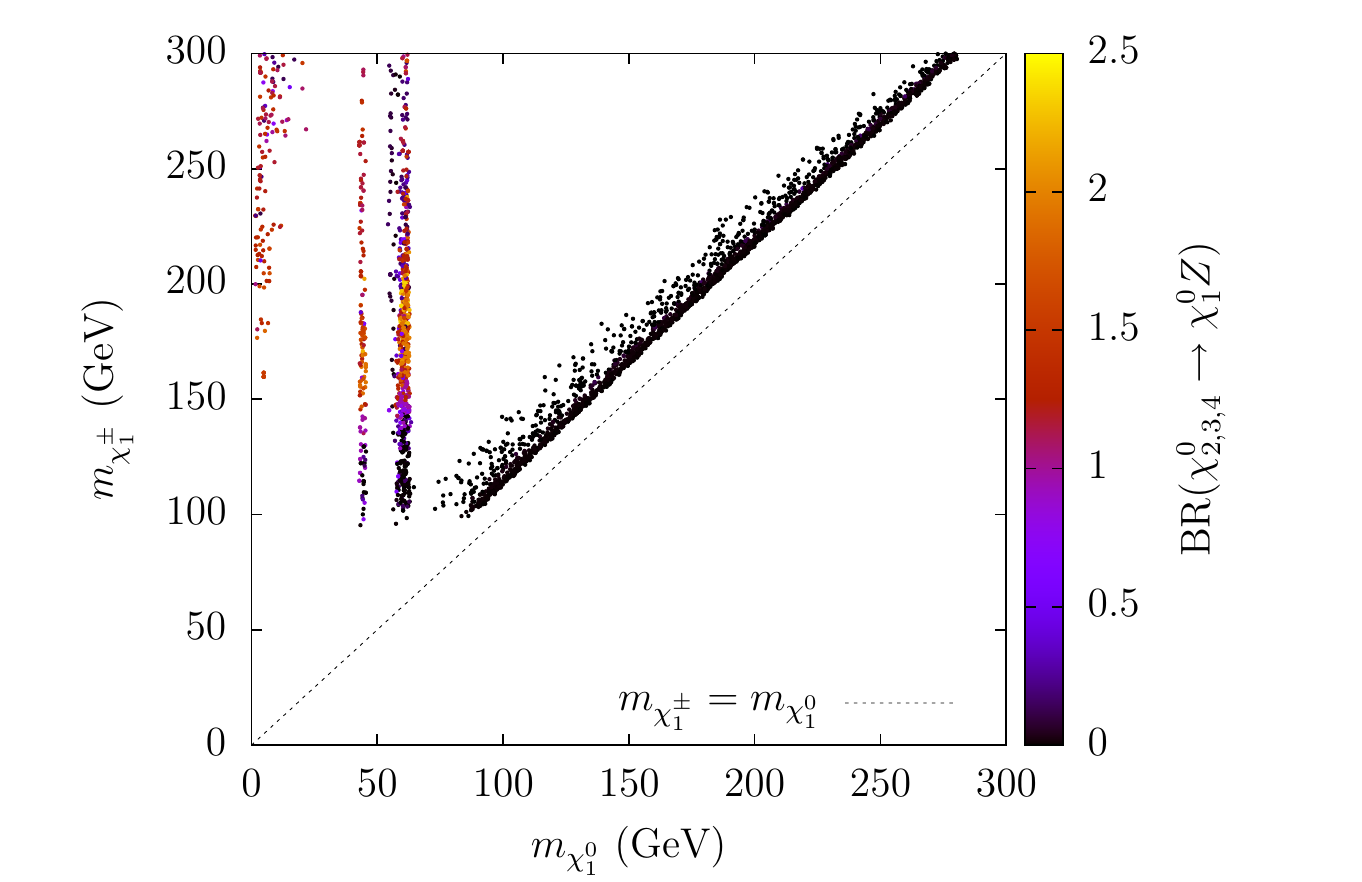}
\vskip 16pt
\includegraphics[height=0.3\textheight, width=0.54\columnwidth ,
clip]{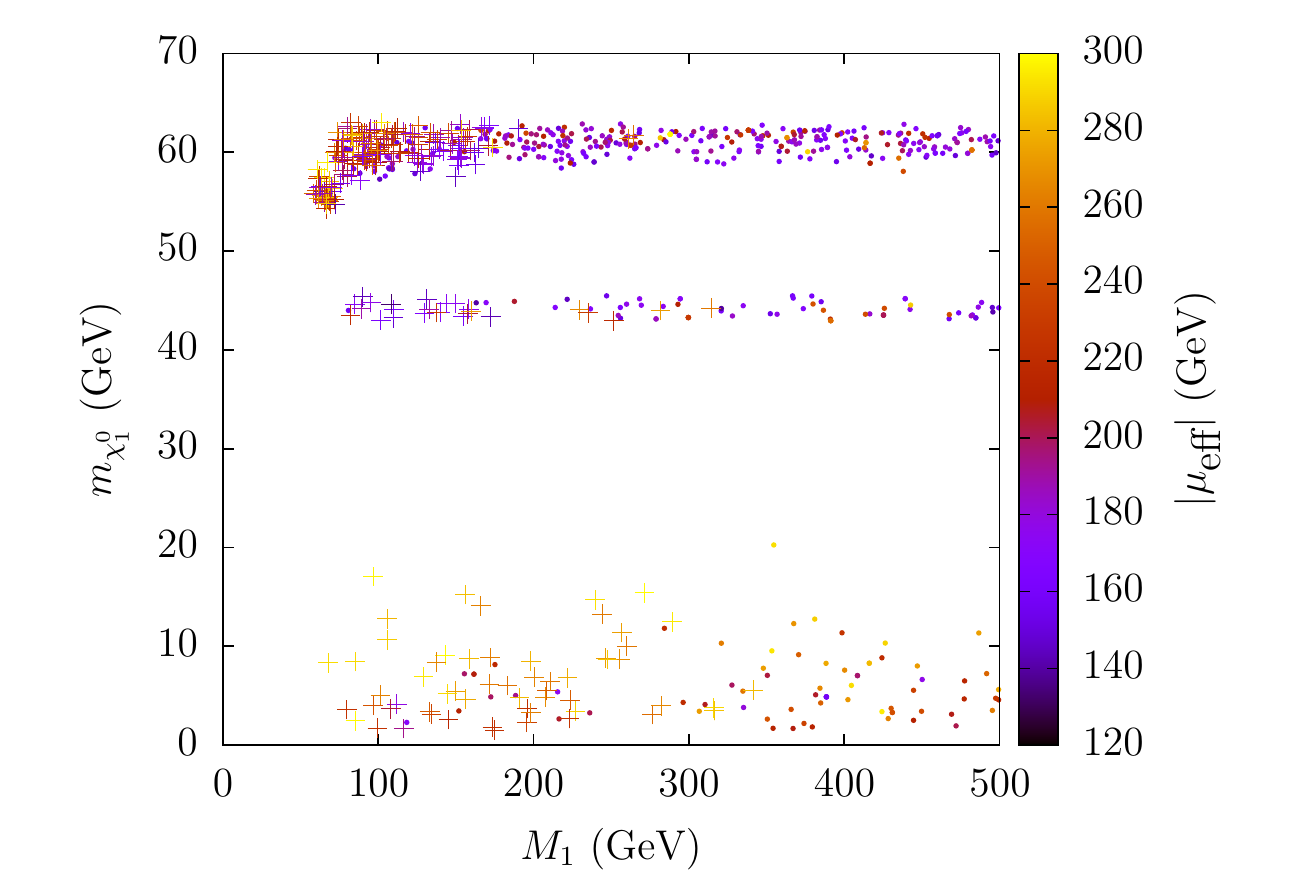}~\hspace{-0.8cm}\includegraphics[height=0.3\textheight, width=0.54\columnwidth ,
clip]{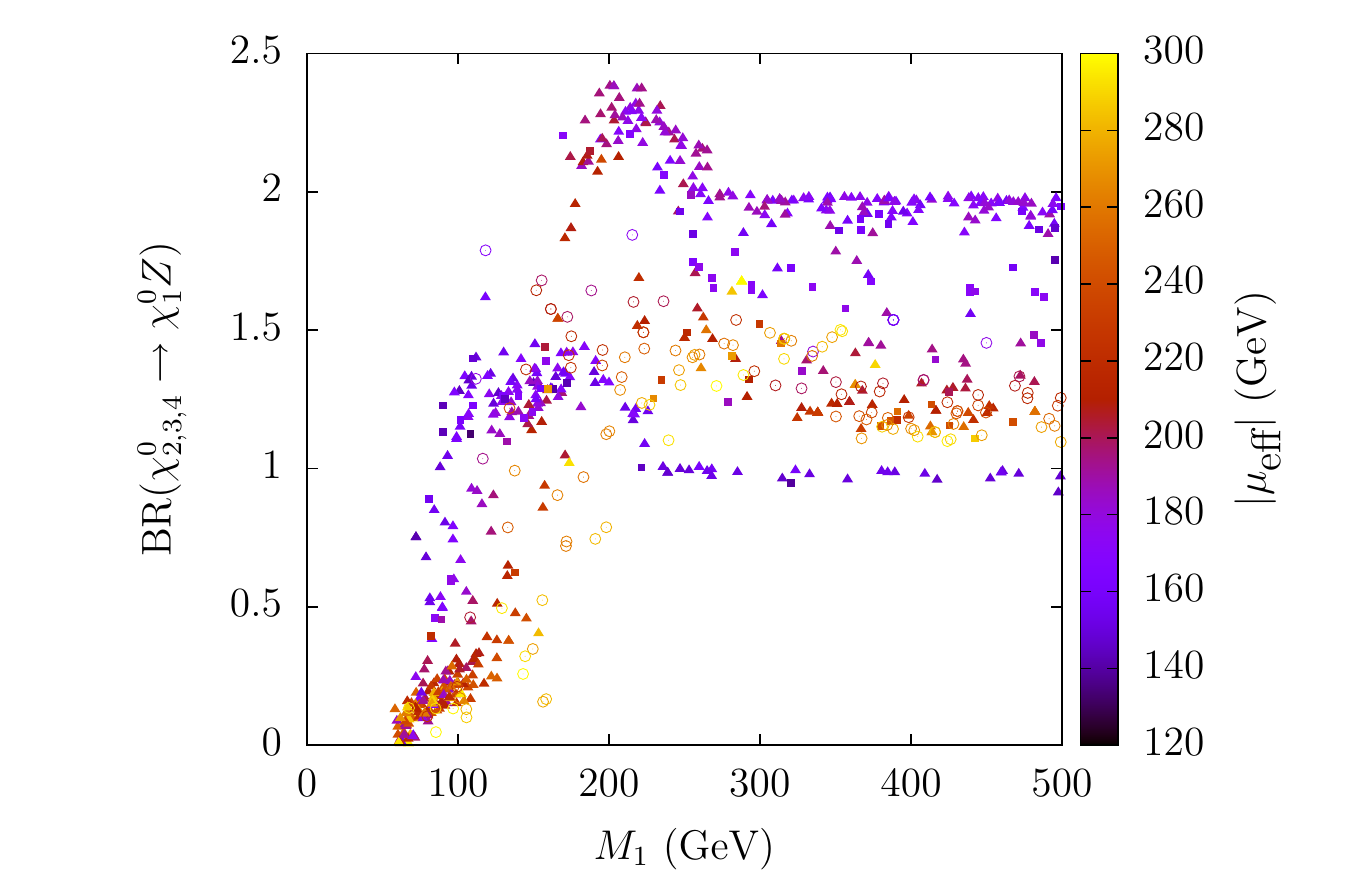}
\caption{Scatter plots in various planes showing variations of relevant
quantities via color-palettes for points in the $Z_3$-symmetric NMSSM parameter
space, with singlino-dominated LSP, obtained via scanning of the same (see text for the ranges used
for various free parameters) and are consistent with all the constraints
discussed in the text including those from the Higgs sector, flavor sector,
DMRD, DMDD-SI and DMDD-SD but before considering the LHC data pertaining to the
ewino sector. See text for details.}
\label{fig:rd-si-sd}
\end{figure}
%
%
\subsection{Impact of bounds from the DM sector}
\label{sec:dm-bounds}
Unless otherwise stated, in the present work, bounds from the DM sector would
imply strict adherence to a relic density within 10\% of the central value of
$\Omega h^2=0.119$ measured by the Planck experiment~\cite{Ade:2015xua, Aghanim:2018eyx}, i.e., $0.107 < \Omega h^2 < 0.131$. Allowed maximum values for the DM-nucleon scattering cross sections are taken
(somewhat conservatively, for DM-mass $\simeq 30$~GeV, for which the
DMDD-SI bound is the strongest) to be
$\sigma^{\rm SI}_{\chi^0_1-p(n)}<4.1\times 10^{-47}$~cm$^2$~\cite{Aprile:2018dbl} and
$\sigma^{\rm SD}_{\chi^0_1-p(n)}<6.3\times 10^{-42}$~cm$^2$~\cite{Aprile:2019dbj}.

In figure~\ref{fig:rd-si-sd} we illustrate various relevant aspects of the
regions of $Z_3$-symmetric NMSSM parameter space that are simultaneously
compatible with all experimental data pertaining to DMRD, DMDD-SI and DMDD-SD.
These aspects are as follows.
\begin{itemize}
\item
The plot on the top is in the $\mntrlone-\mcharone$ plane with the value of
the combined branching fraction of $\ntrltwothreefour$ in the decay mode $\ntrltwothreefour \to \ntrlone Z$ indicated in the palette  which could reach a possible maximum
value of `3'. Visibly, over the dark patch along the diagonal, 
the singlino-dominated DM neutralino is nearly mass-degenerate with the lighter chargino ($\charonepm$) and the next two lighter neutralinos ($\ntrltwothree$) all of which are higgsino-like. 
Hence coannihilation of the DM neutralino with these states is rather efficient. This renders DMRD in the right experimental
ballpark.\footnote{If we assume that the LSP's contribution does not
saturate the observed relic density, we would end up with a somewhat
larger number of allowed points at low LSP mass (thus broadening the funnel
strips) and in the vicinity of the coannihilation region. Note, however, that
the enhanced magnitude of some relevant couplings that result in an increased
DM-annihilation thus leading to such a drop in the thermal relic density could
potentially make the DMDD cross sections breach the experimental constraints.}
It is also important to note that due to this degeneracy, the bounds on the
ewino sector are also much relaxed over this region~\cite{Aaboud:2017leg}.
Hence parameter
points from this region have a good chance to survive bounds obtained from both
DM experiments and the LHC. In fact, ref.~\cite{Cao:2018rix} pointed out
this to be the only region for a singlino-dominated LSP which could exhibit such
a simultaneous compliance with data.
Note that given $\mone < \mueff$ is a possibility in our scan, there may be a situation when $\ntrltwo$ becomes bino-dominated while $\ntrlthreefour$ become higssino-like. For such a spectrum, the decay $\ntrltwo \to \ntrlone Z$
may be kinematically disfavoured while $\ntrlthreefour \to \ntrlone Z$ could open
up and become relevant.

\parindent 20pt
In agreement with ref.~\cite{Cao:2018rix}, our scan also finds strips of
DM-allowed points at LSP masses with the SM Higgs and
$Z$-boson funnels, i.e., for $\mntrlone=\mhtwo/2$ and at $\mntrlone=m_Z/2$,
respectively. However, it appears that these strips extend to much higher
values of $\mcharone$ (apparently limited only by our choice of the upper
limit of $\mueff \, (\lesssim 300$~GeV)) when compared to what was found in
ref.~\cite{Cao:2018rix}. Also, unlike in ref.~\cite{Cao:2018rix}, the 
bottom sections of the funnel strips for the SM Higgs boson and the
$Z$-boson are found to be notably populated. 
We indeed notice that compliance with DMDD-SD data is facilitated with
low values of $\mone$, as discussed in section~\ref{subsec:inter}.
In this region there is a 
substantial mass difference between $\charonepm$ and $\ntrlone$ due to which hard enough leptons are expected from decays of $\charonepm$. Furthermore, 
the 3-body decays
$\ntrltwo,~\ntrlthree \rightarrow \ell \bar{\ell} \ntrlone$
(presumably via an off-shell $Z$ boson)
could contribute significantly. Thus, 
this region is expected to get severely constrained from tri-lepton searches 
at the LHC~\cite{Sirunyan:2017lae, Sirunyan:2018ubx, CMS:2017fdz}. One could as well expect a corresponding 3-body decay of $\ntrltwothree$ that involves a bottom quark pair. Hence searches 
involving $b$-jets in the final states~\cite{Sirunyan:2018ubx} are likely
to get sensitive to the said region of parameter space.

\parindent 20pt
Furthermore, we find a DM-allowed region 
with lighter LSP masses ($\lesssim 20$~GeV) possessing funnels in
light singlet scalars (mostly $\aone$, and only occasionally, $\hone$).
Refs.~\cite{Ellwanger:2016sur, Ellwanger:2018zxt, Cao:2018rix} had
correctly argued on the difficulty in realizing an $\aone$ funnel.
However, as envisaged in ref.~\cite{Ellwanger:2018zxt}, we now find a generic region with a
singlino-dominated LSP with mass $\lesssim 20$~GeV that possesses
$\aone$ funnel for even (an optimally) small `$\lambda$' along with 
rather large $\alambda$. As predicted, the region indeed yields a rather light $\hone$
which, ref.~\cite{Cao:2018rix} argued, would yield too large a
DMDD scattering rate to survive the experimental data. Here, it is our
specific observation that a suitably low value of $\mone$ could again do the trick by pushing the DMDD rate down to a safe 
level.\footnote{Some such situations are discussed in ref.~\cite{Domingo:2018ykx} as specific benchmark points. However, given that the work focuses on the impacts of the
LHC data, it remains agnostic as to whether such points would satisfy
various DM-related constraints but for the DMRD upper bound. We observe that
most of these points possess a rather light $\hone$ ($\mhone \, (\lesssim 20 \, \mathrm{GeV})$) which would make it difficult to survive
DMDD bounds unless for suitable $\mone < \mueff$ thus yielding a bino-like $\ntrltwo$.}
A closer inspection reveals that funnels at work for a singlino-like LSP with
mass $\lesssim 40$~GeV can be that of $\hone$ or
$\aone$ or both, simultaneously. In addition, emergence of points only in discrete strips, even though the LSP and the light scalar masses are varying, are due to stringent
requirement of having the relic density within a specific band about its observed
central value while satisfying the DMDD bounds.

\parindent 20pt
Of some interest are the points in darker shades in the funnel strips.
These are the points for which the collective branching fractions in the
decay modes of $\ntrltwothreefour$ containing a real $Z$-boson are tiny.
Thus, it may be expected that
these could evade some pertinent collider bounds while being still consistent
with all DM data, unless $\mcharone$ is too small, as is the case
 at the bottom of these strips. This is since the latter kinematically
prohibits the decay(s) of one or more of the participating heavier neutralinos
($\ntrltwothreefour$) to $\ntrlone Z$. Nevertheless, three-body decays (via an off-shell $Z$-boson) into leptonic final states may remain significant, as discussed before. 
In addition, we find regular (sparse)
population of darker points within the strips representing $\hsm$
($Z$-boson and $\hone/\aone$) funnel(s) for higher values of $\mcharone$ as well.
These result from opening up of new decay modes involving lighter Higgs bosons for $\ntrltwothreefour$ due to genuine (dynamical) suppressions of the 
strengths for the
$\ntrltwothreefour \ntrlone Z$ interaction in the presence of competing
$\ntrltwothreefour \ntrlone h_i$ interactions. Clearly these points need 
to be subjected to thorough examination to ascertain their viability against 
LHC data. We undertake this exercise, for relevant final states involving 
leptons mostly, using \checkmate in section~\ref{subsec:checkmate} 
with reference to a few benchmark points picked from all the three funnel 
regions. 
%
\item
The plots in the bottom row of figure~\ref{fig:rd-si-sd} convey the interplay
of $\mone$ and $\mueff$ keeping $\mntrlone$ and the combined branching fraction
BR$( \ntrltwothreefour \to \ntrlone Z)$ in reference. Thus, while the
left plot reveals the funnel strips over specific $\mntrlone$ ranges
(thus corresponding exactly to the plot on the top) having
the branching fraction to $Z$-boson either less (indicated by `+' marks) or
greater (indicated by circular blobs) than 1.5, the right plot explicitly
displays the same branching fraction with the three specific (funnel)
ranges for the associated $\mntrlone$ being indicated by three different
symbols: `$\blacktriangle$' for the SM Higgs funnel, `$\blacksquare$' for 
the $Z$-boson funnel and `$\circ$' for the singlet-like scalar(s)
funnel. These two plots clearly reveal that to achieve a dominant
($\geq 1.5$) combined branching fraction to every other mode save $\ntrlone Z$
(thereby evading relevant collider bounds;
represented by the `+' symbol) one requires
$\mone < \mueff$. It is somewhat curious to note that the combined branching fraction to $Z$-boson
could systematically go down to a value of $\approx 1$ but does not drop further
if $\mone > 250$~GeV.
\end{itemize}
%
\begin{figure}[t]
\centering 
\hspace{-1.4cm}\includegraphics[height=0.3\textheight, width=0.57\columnwidth ,
clip]{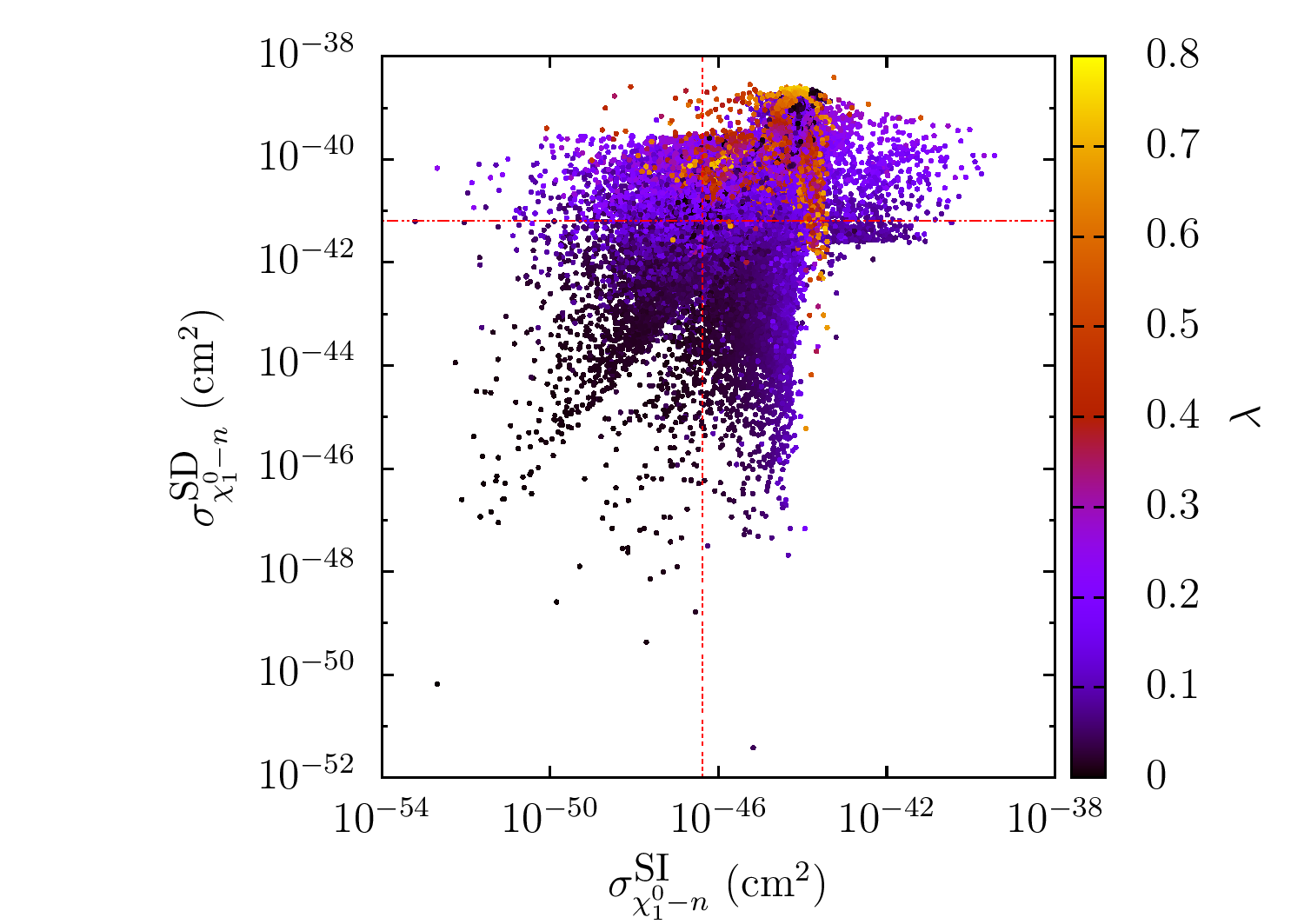}
\hspace{-1.2cm}\includegraphics[height=0.3\textheight, width=0.57\columnwidth ,
clip]{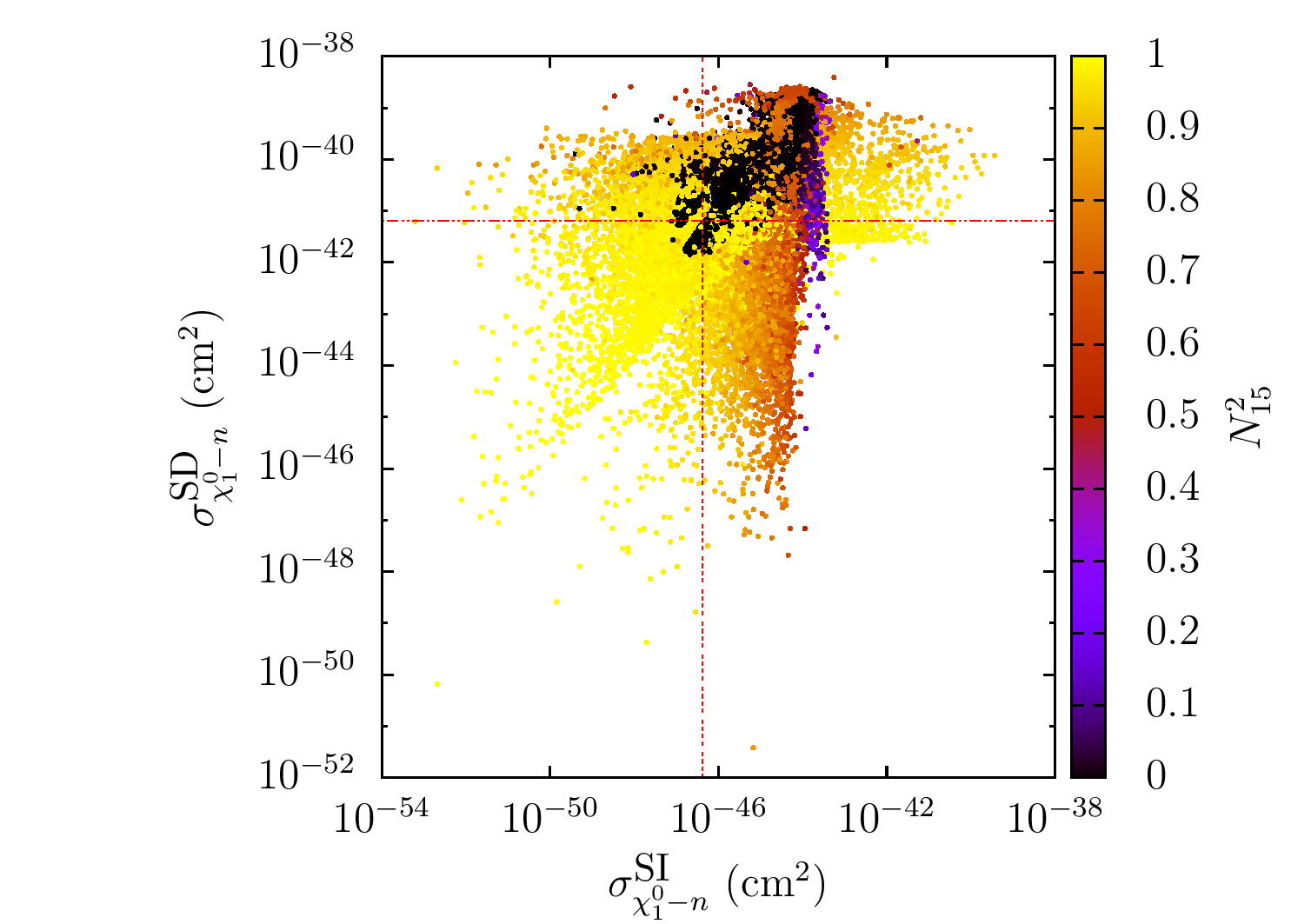}
\caption{Scatter plots in the plane of
$\sigma^{\mathrm{SI}}_{\ntrlone-n}-\sigma^{\mathrm{SD}}_{\ntrlone-n}$ 
indicating the
values of $\lambda$ (singlino fraction in the LSP ($=N_{15}^2$)) from the palette in the left (right) plot while satisfying $0.107 < \Omega h^2 < 0.131$. Observed upper bounds
of the DMDD-SI and DMDD-SD cross sections are indicated by the red-dashed lines
vertical and parallel, respectively, to the abscissa thus indicating that points in
the bottom left quadrant simultaneously satisfy bounds on both.
}
\label{fig:SI-SD-var}
\end{figure}
%

In figure~\ref{fig:SI-SD-var} we illustrate how the values of `$\lambda$'
(left) and hence $N_{15}^2$ (the singlino admixture in the LSP) are
distributed in the
$\sigma^{\mathrm{SI}}_{\ntrlone-n}-\sigma^{\mathrm{SD}}_{\ntrlone-n}$ 
plane. In both plots,
points in the bottom left
quadrant are the only ones that are allowed by both DMDD-SI and DMDD-SD
data. From the left plot one can clearly see that low values of
`$\lambda$' ($\lesssim 0.2$) are preferred. This is not unexpected for the
following reasons.
\begin{itemize}
\item
First, the DMDD-SI
cross section dominantly involves coupling of a DM(LSP) pair to the
singlet-like Higgs
bosons ($\hone \ntrlone \ntrlone$) which is enhanced for a mixed
singlino-higgsino LSP. Given the higgsino admixture in an otherwise 
singlino-dominated LSP is proportional to `$\lambda$'
(for given fixed values of $\mntrlone$ and $\mueff$), the coupling in context grows with its value and could lead to a large enough SI cross section
that is ruled out by the experiments.
\item
Second, the DMDD-SD cross section, in contrast, involves coupling of a 
DM(LSP) pair to a $Z$-boson. This, on the other hand, depends on the higgsino content of an  otherwise singlino-dominated LSP and hence grows as $\lambda^2$
(for given fixed values of $\mntrlone$ and $\mueff$). This could result in a large enough SD cross section which again could be ruled out by relevant experiments.
\end{itemize}

\parindent 20pt
The plot on the right first corroborates the correlation between `$\lambda$' and $N_{15}^2$ that is explained above, i.e., the smaller is the value of
`$\lambda$', the smaller (larger) is the higgsino (singlino) admixture in the
singlino-dominated LSP. Furthermore, one finds that the reddish/purple part on
the right edge of the plot has an enhanced higgsino fraction in the LSP and
hence always gets ruled out by DMDD-SI data. However, DMDD-SD data may still
allow such `$\lambda$' values (in the bottom right quadrant) which is due 
to somewhat smaller sensitivity of SD rates to `$\lambda$', as hinted
above.
In contrast, the black regions are very special
in the sense that these have the LSP which is bino-dominated (when $\mone$ goes below $m_{\tilde{S}} \sim 2 \kappa \vs$ in our scan). The only admixture that
is pertinent here is in the form of higgsinos (since bino does not mix directly
to singlino at the lowest order) and `$\lambda$' is likely to decouple from
DM physics. Thus, a small higgsino admixture in the LSP could suffice to
result in a large enough SI and SD cross sections that are ruled out by
experiments. Nonetheless, we find a tiny bino-dominated region in the intersection
of the two boundaries that separate the DD-allowed regions. For clarity, it may be mentioned that the points that appear in the allowed (bottom left) quadrant
comply with all DM data and hence are the same data-points that show up in
figure~\ref{fig:rd-si-sd}.
%
\subsection{Benchmark scenarios}
\label{subsec:benchmark}
In this subsection we briefly discuss our strategy to choose a few
representative benchmark points that worth thorough scrutiny against recent LHC
data in order to establish their viability. We choose our benchmark points from
the scan described earlier by ensuring that these all have a singlino-dominated ($> 95\%$) LSP, have low values of $\mueff$ and satisfy basic experimental
constraints mentioned earlier including those from the DM-sector. The scenarios
are divided into three
categories according to the DM-annihilation funnels at work, i.e., singlet
(pseudo)scalar funnel, $Z$-boson funnel and SM-like Higgs funnel. 
Note that we have ensured, apart from satisfying the DMRD and the DMDD
constraints, 
our benchmark points also satisfy various other constraints from indirect DM searches~\cite{Fermi-LAT:2016uux,Giesen:2015ufa,Ibarra:2013zia} thanks to a small 
annihilation cross-section at late times ($\langle \sigma v \rangle \lesssim 
\mathcal{O}(10^{-29}) \, {\rm cm}^3\, {\rm s}^{-1})$.
Next, we look for
if the combined decay branching fraction of the heavier neutralinos to
$\ntrlone Z$ could be on the smaller side so that such points stand higher 
chance of evading LHC constraints on the lighter ewino sector. Furthermore, we
try to ensure that the decay branching fraction for $\charonepm \to  \ntrltwo W^\pm$ competes or even exceeds that for $\charonepm \to  \ntrlone W^\pm$
adopted in the standard paradigm for experimental analyzes. This would further relax the existing bounds in this sector.

In table~\ref{tab:BPs} we present these benchmark points by
indicating the relevant input parameters, the resulting spectra, the contents of
the LSP and the next-to-lightest neutralinos, various relevant branching fractions along
with the values for the DM observables. Finally, we summarize for each of these
points, their status in view of recent LHC analyzes obtained via {\tt CheckMATE}.  
%
\afterpage{
\begin{table}[H]
\begin{center}
{\small\fontsize{7.5}{7.5}\selectfont{
\begin{tabular}{|c|@{\hspace{0.06cm}}c@{\hspace{0.06cm}}|@{\hspace{0.06cm}}c@{\hspace{0.06cm}}|@{\hspace{0.06cm}}c@{\hspace{0.06cm}}|c|@{\hspace{0.06cm}}c@{\hspace{0.06cm}}|}
\hline
\Tstrut
 & \makecell{Singlet (pseudo)scalar \\ funnel} & \multicolumn{2}{c|}{\makecell{$Z$-boson \\ funnel}}& \multicolumn{2}{c|}{\makecell{SM-like Higgs\\ funnel}}\\
\hline
\Tstrut
$\lambda$  &  $8.72\times 10^{-2}$  &  $0.181$         &  $0.133$          &  $0.120$  & $0.160$   \\
$\kappa$   &  $2.43\times 10^{-3}$  &  $-1.28\times 10^{-2}$ & $1.23\times 10^{-2}$ &  $1.74\times 10^{-2}$  & $1.76\times 10^{-2}$   \\
$\tan\beta$        &  33.69    &  26.56   & 11.86    &  39.61  & 9.13 \\
$A_\lambda$~(TeV)  &  10.15    &  7.67    & 2.56     &  8.90   & 2.81  \\
$A_\kappa$~(GeV)   &  $-58.25$   &  51.42   & $-13.93$   &  $-35.90$ & $-0.52$   \\
$\mu$~~(GeV)       &  297.65   &  297.81  & 230.46   &  193.10 & 250.63   \\
$M_1$~(GeV)        &  96.85    &  97.91   & 137.64   &  115.00 & 87.10  \\
$M_2$~(GeV)        &  485.83   &  689.15  & 556.26   &  575.12 & 417.42   \\[0.05cm]
\hline
\Tstrut
$m_{\chi_1^0}$~(GeV)    &  17.07   &  43.40   & 43.78   &  57.40  & 55.49   \\
$m_{\chi_2^0}$~(GeV)    &  94.00   &  95.03   & 129.05  &  107.26 & 83.26  \\
$m_{\chi_3^0}$~(GeV)    &  298.79  &  306.86  & 240.02  &  204.84 & 247.11   \\
$m_{\chi_4^0}$~(GeV)    &  314.69  &  315.71  & 245.32  &  208.28 & 265.15  \\
$m_{\chi_5^0}$~(GeV)    &  543.61  &  749.64  & 611.46  &  631.06 & 468.50  \\

$m_{\chi_1^\pm}$~(GeV)  &  297.37  &  303.73  & 231.96  &  196.67 & 242.56  \\
$m_{\chi_2^\pm}$~(GeV)  &  543.68  &  749.66  & 611.47  &  631.08 & 468.51  \\
$m_{h_1}$~(GeV)         &  8.49    &  41.11   & 40.68   &  48.17  & 52.62  \\
$m_{h_2}$~(GeV)         &  125.53  &  125.54  & 124.75  &  125.65 & 122.90  \\
$m_{a_1}$~(GeV)         &  37.65   &  56.25   & 34.23   &  55.12  & 20.47  \\[0.05cm]
\hline
\Tstrut
$N_{11}$, $N_{21}$  & $0.03,~~0.98$   & $0.03,~~0.98$  & $0.05,~~0.95$   & $0.09,-0.93 $  &  $0.17,-0.96 $  \\
$N_{12}$, $N_{22}$  & $\!\!\!\!-0.01,-0.01$   & $\!\!\!\!-0.01,-0.01$  & $\!\!\!\!-0.02,-0.03$   & $\!\!\!\!-0.02,~~0.02$  &  $\!\!\!\!-0.03,~~0.02$  \\
$N_{13}$, $N_{23}$  & $0.01,~~0.15$   & $\!\!\!\!-0.01,~~0.15$  & $0.02,~~0.26$   & $0.06,-0.28 $  &  $0.05,-0.18 $ \\
$N_{14}$, $N_{24}$  & $\!\!\!\!-0.05,-0.05$   & $\!\!\!\!-0.10,-0.01$  & $\!\!\!\!-0.10,-0.15$   & $\!\!\!\!-0.12,~~0.14$  &  $\!\!\!\!-0.12,~~0.06$  \\
$N_{15}$, $N_{25}$  & $0.99,-0.03$   & $0.99,~~0.00$  & $0.99,-0.07$   & $0.98,~~0.12 $  &  $0.97,~~0.18$  \\[0.05cm]
\hline
\Tstrut
BR($\chi^\pm_1\to\chi_1^0 W^\pm $)  &  0.13  &  0.37  & 0.47  &  0.59& 0.39  \\[0.05cm]
BR($\chi^\pm_1\to\chi_2^0 W^\pm $)  &  0.87  &  0.63  & 0.53  &  0.41& 0.61  \\[0.05cm]
\hline
\Tstrut
BR($\chi^0_2\to\chi_1^0 Z $)    &  0.00  &  0.00  & 0.00  &  0.00 & 0.00  \\[0.05cm]
BR($\chi^0_2\to\chi_1^0 h_1 $)  &  0.92  &  1.00  & 0.95  &  1.00 & 0.00  \\[0.05cm]
BR($\chi^0_2\to\chi_1^0 h_2 $)  &  0.00  &  0.00  & 0.00  &  0.00 & 0.00  \\[0.05cm]
BR($\chi^0_2\to\chi_1^0 a_1 $)  &  0.08  &  0.00  & 0.03  &  0.00 & 1.00  \\[0.05cm]
\hline
\Tstrut
BR($\chi^0_3\to\chi_1^0 Z $)    &  0.04  &  0.22  & 0.25  &  0.18 & 0.06  \\[0.05cm]
BR($\chi^0_3\to\chi_2^0 Z $)    &  0.25  &  0.19  & 0.24  &  0.33 & 0.22  \\[0.05cm]
BR($\chi^0_3\to\chi_1^0 h_1 $)  &  0.00  &  0.01  & 0.02  &  0.00 & 0.01  \\[0.05cm]
BR($\chi^0_3\to\chi_2^0 h_1 $)  &  0.01  &  0.03  & 0.16  &  0.07 & 0.01  \\[0.05cm]
BR($\chi^0_3\to\chi_1^0 h_2 $)  &  0.07  &  0.10  & 0.33  &  0.41 & 0.29  \\[0.05cm]
BR($\chi^0_3\to\chi_2^0 h_2 $)  &  0.63  &  0.45  & 0.00  &  0.00 & 0.41  \\[0.05cm]
BR($\chi^0_3\to\chi_1^0 a_1 $)  &  0.00  &  0.00  & 0.00  &  0.01 & 0.00  \\[0.05cm]
\hline
\Tstrut
BR($\chi^0_4\to\chi_1^0 Z $)    &  0.09  &  0.18  & 0.38  &  0.67 & 0.36  \\[0.05cm]
BR($\chi^0_4\to\chi_2^0 Z $)    &  0.74  &  0.54  & 0.54  &  0.30 & 0.52  \\[0.05cm]
BR($\chi^0_4\to\chi_1^0 h_1 $)  &  0.00  &  0.01  & 0.00  &  0.00 & 0.00  \\[0.05cm]
BR($\chi^0_4\to\chi_2^0 h_1 $)  &  0.00  &  0.00  & 0.00  &  0.00 & 0.00  \\[0.05cm]
BR($\chi^0_4\to\chi_1^0 h_2 $)  &  0.03  &  0.17  & 0.07  &  0.01 & 0.03  \\[0.05cm]
BR($\chi^0_4\to\chi_2^0 h_2 $)  &  0.14  &  0.10  & 0.00  &  0.00 & 0.08  \\[0.05cm]
BR($\chi^0_4\to\chi_2^0 a_1 $)  &  0.00  &  0.00  & 0.01  &  0.02 & 0.01  \\[0.05cm]
BR($\chi^0_4\to\chi_2^0 a_2 $)  &  0.00  &  0.00  & 0.00  &  0.00 & 0.00  \\[0.05cm]
\hline
\Tstrut
$\Omega h^2$  &  0.12447  &  0.12739  & 0.12713 &  0.13002 & 0.10723  \\[0.10cm]
$\sigma^{\rm SI}_{\chi^0_1-p(n)}$~(cm$^2$)  &    $0.8(1.1)\times 10^{-47}$&  $4.3(4.0)\times 10^{-47}$  & $2.5(2.3)\times 10^{-47}$ &  $6.6(8.4)\times 10^{-48}$ & $4.7(5.1)\times 10^{-47}$   \\[0.30cm]
$\sigma^{\rm SD}_{\chi^0_1-p(n)}$~(cm$^2$)  &    $2.3(1.7)\times 10^{-43}$&  $4.6(3.5)\times 10^{-42}$  & $3.8(2.9)\times 10^{-42}$ &  $4.9(3.8)\times 10^{-42}$ & $5.8(4.4)\times 10^{-42}$   \\[0.15cm]
\hline
\Tstrut
\texttt{CheckMATE} result  &    Allowed   &  Allowed  &  Allowed  & Allowed& Allowed  \\
$r$-value               &   0.97 &  0.57     &  0.81  & 0.70& 0.90  \\
Analysis          &   CMS$\_$SUS$\_$16$\_$039~\cite{Sirunyan:2017lae} &  CMS$\_$SUS$\_$16$\_$039  &  CMS$\_$SUS$\_$16$\_$039  & CMS$\_$SUS$\_$16$\_$039& CMS$\_$SUS$\_$16$\_$039\\
Signal region&    SR$\_$G05 &  SR$\_$A30  &  SR$\_$A30  & SR$\_$A25& SR$\_$A30\\[0.05cm]
\hline
\end{tabular}
}}
\caption{Benchmark points in the $Z_3$-symmetric NMSSM parameter space offering
different annihilation-funnels for a singlino-dominated DM neutralino along with
resulting spectra, compositions of the DM and the next-to-lightest neutralinos, various
relevant branching fractions of the light ewinos, the values of the DM relic
density, the DMDD-SI and the DMDD-SD cross sections. Also indicated are the
`$r$'-values (see section~\ref{subsec:checkmate}) returned by \checkmate along
with the reference LHC analyzes and the most sensitive Signal Regions (SR).
{\tt NMSSMTools}-{\tt v5.3.0} is used to generate the spectra and to calculate the
decay branching fractions of various ewinos. The DM-observables are estimated
with the \micromegas ~package built-in in \nmssmtools. The fixed values of various soft parameters used are as follows:
$\mthree=m_{{(\tilde{Q}, \tilde{U}, \tilde{D})_{1,2}}}
        =m_{\tilde{L}, \tilde{E}} = A_{b,t}= 5$~TeV,
$m_{{(\tilde{Q}, \tilde{U})_3}} = 5.5$~TeV and $A_\tau=5.6$~TeV.
}
\label{tab:BPs}
\end{center}
\end{table}
}
%
%
\subsection{Impact of recent LHC results: a \checkmate\!\!-based analysis}
\label{subsec:checkmate}
In this section we describe the status of the benchmark scenarios presented in
table~\ref{tab:BPs} in the light of the LHC results. As can be seen, these
scenarios feature a light higgsino-like chargino, $\charonepm$ and several light
neutralinos, $\ntrli$, with $i \in \{1,3\}$ when only singlino- and higgsino-like
states are considered and $i \in  \{1,4\}$ when, in addition, a light bino-like
state is allowed.

It may be reiterated that when we consider only singlino- and higgsino-like
light neutralinos in the presence of a light (pseudo-) scalar Higgs in the
spectrum, the following decay channels are of importance:
$$ \charonepm \rightarrow \ntrlone  W^{\pm}, 
~ \ntrli \rightarrow \ntrlone Z/h/a,   \quad (i=2,3)$$ where
$h \equiv \{\hone, \htwo (\hsm)\}$ (`$a$') represents a
$CP$-even ($CP$-odd) Higgs boson. In the presence of a bino-like state in the spectrum,
typically $ \ntrltwo$ for our benchmark points, the following additional decay modes can be relevant too:
\[ \charonepm \rightarrow \ntrltwo W^{\pm},~
\ntrltwo \rightarrow \ntrlone h/a, 
~\ntrli \rightarrow \ntrltwo Z/h/a,  \quad (i=3,4). \]
Depending on the mass-difference between the heavier higgsino-like states and
$\ntrlone$, on- or off-shell gauge/scalar bosons may appear in the above
decays of the light ewinos. Since we mainly focus on the uncompressed region,
with rather sizable mass-split between the heavier higgsino-like states and
$\ntrlone$, on-shell gauge bosons feature in all our benchmark scenarios. 

Considering final states with leptons, the following final states are going to
be relevant.
\begin{itemize} 
\item Chargino pair production ($pp \to \chi_1^\pm \chi_1^\mp$) can lead to
$2\ell + \slashed{E}_T$  (missing $E_T$ or MET). In the
presence of a bino-like $\ntrltwo$, there could be significant number of events
with up to four accompanying $b$-jets, assuming the Higgs boson in the cascade
dominantly decays into two $b$-quarks.
\item Chargino-heavier neutralinos associated production ($pp \to \chi_1^\pm
\ntrltwothreefour$) can lead to $3 \ell+\slashed{E}_T$ and $\ell + 2b +
\slashed{E}_T$. As in the previous	case, the presence of a bino-like $\ntrltwo$,
either produced in the hard scattering or in the cascade of heavier neutralinos,
might lead to final states with an enhanced $b$-jet multiplicity.
\item Finally, heavier neutralino pair production could lead to up to $2\ell
+2$-jets/$4\ell +\slashed{E}_T$ where, in our case, the pairs of leptons come from
the decay of on-shell $Z$-bosons. The presence of a bino-like $\ntrltwo$ in the
cascade, as before, would ensure enhanced $b$-jet multiplicity in the final
state. 
\end{itemize}	

We use \texttt{CheckMATE}-{\tt v2.0.26} to test our benchmark scenarios against relevant
experimental analyzes by the ATLAS and the CMS collaborations (which are already
implemented in \checkmate and have been validated) at the 13~TeV LHC with up to
36~\fbinv ~worth data.
%
%
The mono-jet/$\gamma + \slashed{E}_T$~\cite{Aaboud:2018doq,Aaboud:2017phn}
are relevant for pair production of the lightest neutralino, together with an
ISR jet or a photon. Searches for
two soft leptons~\cite{Sirunyan:2018iwl, Aaboud:2017leg} 
can be relevant for compressed spectra of light ewinos. While searches in these
final states could, in general, put reasonable constraints on the
chargino-neutralino spectra, these are not expected to be much constraining in
the current context. In the case of mono-jet/mono-photon searches, the
insensitivity stems from the small production cross-section of $\ntrlone$ pair
in the present scenario. Soft lepton searches are insensitive since in our case
the heavier ewinos and the lightest neutralino are already well-separated in mass.

Several other searches for strongly interacting particles have been performed by
both the ATLAS and the CMS collaborations. The inclusion of $b$-tagged jets,
together with leptons can be relevant in our present context. However, these
searches consider large jet multiplicity (typically $\geq 4-6$ jets).
Generic absence of large jet multiplicity in our situations make them immune to
any constraint whatsoever derived from these searches.  

The most relevant searches, in our case, involve multi-leptons and $b$-tagged
jets along with $\slashed{E}_T$, low jet multiplicity~\cite{Sirunyan:2017lae, CMS:2017fdz, Aaboud:2017dmy}.\footnote{Final states
involving `$\tau$' leptons have also been considered in the literature~\cite{Aaboud:2017nhr,Sirunyan:2017lae}. However, our benchmark scenarios are not
sensitive to the signal regions discussed in those works.}
Out of the multi-lepton analyzes implemented in the \checkmate version that we
employed, the most stringent constraints appear to arise from the
$3\ell+\etmiss$ final states, as well as from the ones with an
opposite-sign di-lepton pair in the final states~\cite{Sirunyan:2017lae, Sirunyan:2018ubx, CMS:2017fdz}.\footnote{Final states
with leptons and $b$-jets have been considered in ref.~\cite{Sirunyan:2018ubx} and certain signal regions discussed there can be
relevant for our present study. However, the experimental results have not been
implemented in \checkmate version we used and hence it is beyond the scope of
the present work.}

We use \texttt{MadGraph5}-{\tt v2.4.3}~\cite{Alwall:2014hca} to simulate ewino
pair/associated production. Events are generated for $p p \rightarrow \chi_j \chi_k$, ($\chi_j \in \{\chi_i^0, \chi_1^{\pm}\}$), with up to
one additional parton in the final state. These result in 10 (15) distinct
production channels when 3 (4) light neutralino states are considered. For each
production channel, 0.3 million parton level events are generated.
We then use the built-in version of \texttt{PYTHIA6}~\cite{Sjostrand:2006za, Sjostrand:2007gs} for showering and hadronization and
for decays of unstable particles. We have used the \texttt{MLM}~\cite{mlm, Mangano:2006rw} prescription for the matching of jets from matrix elements with those from parton showers, as implemented in \madgraph.

Typically, on merging and matching of partonic jets, the number of
simulated events per production channel reduces to around 0.2 million, on an
average. Such a volume of generated event-samples is expected to be healthy 
enough to ensure a stable statistics and hence could be used for reliable 
estimates in subsequent analyzes. The cross sections for all the processes have 
been computed at the leading order in \texttt{MadGraph}.
A flat $K$-factor of 1.25~\cite{Fiaschi:2018hgm} has
been multiplied to the cross sections of all relevant ewino pair
production processes to factor in the approximate NLO+NLL 
contributions. This is expected to help \texttt{CheckMATE} make 
conservative estimates of the lowest values of the ewino masses that the
recent LHC data could allow. Finally, we have used \checkmate\cite{Dercks:2016npn} (see also~\cite{Read:2002hq, Cacciari:2008gp, Cacciari:2005hq, Cacciari:2011ma, deFavereau:2013fsa}) to examine the viability of 
the benchmark scenarios in the light of 13~TeV LHC results. \checkmate reports
an $r$-value for each of the benchmark scenarios where
$r = (S-1.64 \Delta S)/ S95$ and `$S$', $\Delta S$ and $S95$ denote the
predicted number of signal events, its Monte Carlo error and the experimental
limit on `$S$' at 95\% confidence level, respectively. 

The benchmark scenarios in table~\ref{tab:BPs} are so chosen that they
yield $r<1$ which, going by the \checkmate convention, are dubbed `allowed'
by the LHC analyzes employed for the purpose. We are aware of a stricter
criteria used in some literature (say, $r<0.67$~\cite{Domingo:2018ykx}) for
definiteness in such a conclusion. In that sense, our approach is only
semi-conservative. Thus, the best that can be said about these points is that
most of them are on the verge of being ruled out by the LHC experiments and
might soon get to be so with some additional data. However, at present, they
are indicative of how low a $\mueff$ could still be viable under different
scenarios when the LSP is singlino-dominated. Table~\ref{tab:BPs} reveals
that $\mueff$ as low as $\sim 200$~GeV cannot yet be ruled out with a
reasonable certainty.

Recently the ATLAS collaboration has analyzed 139~$\rm{fb}^{-1}$ of data and has derived constraints by studying the pair production of
charginos where they used the di-lepton $\!\!+\slashed{E}_T$ data for the purpose~\cite{Gignac:2668823}. However, the analysis assumes that the chargino decays 100 \% of the times to $\ntrlone W^{\pm}$. Since in the presence of a bino-like 
$\ntrltwo$, as demonstrated in our benchmark scenarios, there is a substantial contribution from the decay mode $\charonepm \rightarrow \ntrltwo W^{\pm}$, the constraints derived from the above analysis do not apply
directly to our cases.

Further, we observe that inclusion of electroweak productions of heavier neutralinos ($\chi^0_{3,4}$, which though are not too heavy in the absolute sense) seems to play an important role in further exclusion of the $\mcharone-\mntrlone$ plane (see table~\ref{tab:BPs}) beyond what is reported in the literature, although the possibilities find a mention there~\cite{Ellwanger:2018zxt}. This is since these additional modes contribute to the final states that are instrumental in the exclusion.
\section{Conclusions}
\label{sec:conclusions}
A low value of $\mueff$ is known to ensure an enhanced degree of `naturalness'
in a $Z_3$-symmetric NMSSM scenario. An interesting possibility in such a scenario is a light singlino-dominated LSP DM. These two together form the edifice of a singlino-higgsino LSP as a possible candidate for the DM.
Motivated by these, in this work, we have explored in some detail the viability of relatively low values of $\mueff$ with
the LSP being singlino-dominated.

We agree with the observations made in the recent literature that for a
singlino-dominated LSP it is not easy to meet the relevant constraints from the
DM and the collider sectors simultaneously. Compliance has been reported only
when the higgsino-like ewinos are nearly degenerate with the singlino-like LSP.
This ensures its efficient coannihilation with a degenerate
higgsino-like state thus producing a relic at the right (experimentally
observed) ballpark. At colliders, this presents a compressed spectrum that
results in relaxed bounds on the higgsino-like states which could then be light
and still evading generic searches.

We have presented a rigorous analysis of regions of the target parameter space (with
relatively light singlino-like LSP of mass $\lesssim m_{\hsm}/2$, with a purity level
$> 95\%$ and with relatively
small $\mueff$) which exhibit such an overall compliance with experimental
data. These comprise of theoretically much-motivated regions that offer
DM-annihilation funnels in the SM-like Higgs boson, in the $Z$-boson and in the
singlet-like scalars. The higgsino admixture in the LSP DM is anyway necessary
to secure their optimal annihilation in order to find compliance with the
observed relic density. However, this needs moderation since otherwise the
cross section
for DM scattering off the nucleon in the DMDD experiments (in particular,
DMDD-SD) becomes too large and violates the reported bounds.

We have demonstrated that
allowing for a smaller value of $\mone$ and/or $\mtwo$ ($\sim \mueff$) can be
a helpful one-shot manoeuvre that could favorably tweak the dynamics and
the kinematics simultaneously. This way it helps achieve the right balance among
various relevant interaction strengths and decay branching fractions thus
offering simultaneous compliance with data from both DM experiments and the
colliders. In the process, productions and decays of heavier neutralinos
($\ntrlthree$ and $\ntrlfour$) become relevant and these influence the bounds
that can be obtained on the parameter space from the experimental analyses. In a sense, this presents the scope and the requirement of an
indispensable and nontrivial tempering of the singlino-like LSP for the purpose. Further studies in the area of  tempered
neutralinos in the NMSSM  are in progress~\cite{wip}.
%
\acknowledgments
WA would like to acknowledge the support in the form of funding
available from the Department of Atomic Energy, Government of India for the
Neutrino Project and the Regional Centre for Accelerator-based Particle Physics
(RECAPP) at Harish-Chandra Research Institute (HRI). He also acknowledges the
Indian Statistical Institute, Kolkata for hosting him on a collaborative visit where much of the current work is done. AC acknowledges support from  
the Department of Science and Technology, India, through INSPIRE faculty fellowship, (grantno: IFA 15 PH-130, DST/INSPIRE/04/2015/000110).
He thanks HRI for hosting him during his visit to the place when a
significant progress in the present work was made. AD acknowledges the 
hospitality accorded to him by the School of Physical Sciences, Indian
Association for the Cultivation of Science, Kolkata on his extended visits from
where he continued to work on the present project. The authors
are especially thankful to J. Beuria for his enthusiastic participation in the
initial phase of the collaboration.
They acknowledge the
use of the High Performance Computing facility at HRI.
%
%

%
\end{document}